\documentclass{aa}
\usepackage{graphicx,multirow,natbib,txfonts,url}
\bibpunct{(}{)}{;}{a}{}{,}
\newcommand{\eg}{{\it e.g.,}\/}
\newcommand{\Ha}{{H$\alpha$}}

\begin{document}

\title{EUV lines observed with EIS/Hinode in  a solar prominence}
\author{N. Labrosse\inst{1} \and B. Schmieder\inst{2} \and P. Heinzel\inst{3,2} \and T. Watanabe\inst{4}} \offprints{N. Labrosse}
\institute{SUPA, School of Physics and Astronomy, University of Glasgow, Glasgow G12 8QQ, Scotland\\
\email{Nicolas.Labrosse@glasgow.ac.uk}
\and LESIA, Observatoire de Paris -- Meudon, 92195 Meudon Cedex, France
\and Astronomical Institute, Academy of Sciences, 251 65 Ond\v{r}ejov, Czech Republic
\and National Astronomical Observatory of Japan, Mitaka, Japan}
\date{Received / Accepted}

\abstract 
{{During a multi-wavelength observation campaign  with Hinode and ground-based instruments, a solar prominence has been observed for three consecutive days as it was crossing the West limb in April 2007.}}
{{We report on observations obtained on 26 April 2007 using EIS (Extreme Ultraviolet Imaging Spectrometer) on Hinode. They are analysed in order to give a qualitative diagnostic of the plasma in different parts of the prominence.}} 
{{After correcting for instrumental effects, the rasters  in different wavelengths are presented. Several regions within the same prominence are identified for further analysis. Selected profiles for lines with formation temperatures between $\log(T)=4.7$ and $\log(T)=6.3$, as well as their integrated intensities, are given. The profiles of coronal, transition region, and \ion{He}{ii} lines are discussed. We pay special  attention to the \ion{He}{ii} line which is blended with coronal lines.
}}
{Some quantitative results are obtained by analysing the line profiles. They confirm that depression in EUV lines can be interpreted by  two  mechanisms: absorption of coronal radiation by the hydrogen and neutral helium resonance continua, and emissivity blocking.
{We present {estimates of} the \ion{He}{ii} line integrated intensity in different parts of the prominence according to different scenarios for the {relative contribution of absorption and emissivity blocking on the coronal lines blended with the \ion{He}{ii} line}. We estimate the contribution of the \ion{He}{ii} 256.32~\AA\ line in the \ion{He}{ii} raster image to vary between $\sim$ 44\% and 70\% of the raster's total intensity in the prominence according to the different models used to  take into account the blending coronal lines. The inferred integrated intensities of the \ion{He}{ii} 256~\AA\ line are consistent with theoretical intensities obtained with previous 1D  non-LTE radiative transfer calculations, yielding a preliminary estimate for the central temperature of 8700~K, central pressure of 0.33~dyn cm$^{-2}$, and column mass of $2.5 \times 10^{-4}$~g cm$^{-2}$. 
{The corresponding theoretical hydrogen column density ($10^{20}$~cm$^{-2}$) is about two orders of magnitude higher than those inferred from the opacity estimates at 195~\AA.}
{The non-LTE calculations indicate that the \ion{He}{ii} 256.32~\AA} line is essentially formed in the prominence-to-corona transition region by resonant scattering of the incident radiation.}}
{}

\keywords{Sun: filaments, prominences -- Line: profiles -- Sun: corona -- Sun: UV radiation -- Radiative transfer}

\maketitle

\section{Introduction}

Solar quiescent prominences are cool structures ($10^{4}$~K) embedded in the hot corona ($10^{6}$~K). Their formation and their long life time, up to a few rotations, are still not well understood.
Prominences can be formed by injection of chromospheric plasma or by
condensation of coronal plasma in coronal structures \citep{2010SSRv..151..333M}. Understanding
the dynamics and the thermodynamic properties of the prominence plasma is important to understand the different stages in the prominence life.
Spectroscopy is a very powerful tool, and combined with high spatial
and temporal resolutions we get information on both the fine
structure and dynamics
\citep[see][for a review of spectral diagnostics and non-LTE modelling techniques]{2010SSRv..151..243L}.

In the spectral range of the EIS instrument \citep[Extreme ultraviolet Imaging Spectrometer, described in][]{2007SoPh..243...19C}, several coronal and transition-region (TR) lines are detected \citep[see][]{2007PASJ...59S.857Y}. In this paper we show the first EIS spectra
obtained in a prominence  in this wavelength range and discuss their
signatures. There are basically three categories of lines detected
by EIS within the prominence: coronal lines, transition-region lines
and the cool \ion{He}{ii} 256~\AA\ resonance line.  Previous analysis of such lines in presence of prominences has been done by using mostly filtergrams  or integrated intensity rasters \citep[e.g. using CDS, the Coronal Diagnostics Spectrometer,][]{1995SoPh..162..233H}. Prominence material in the corona may absorb and block the coronal emission and emit transition-region radiation.

 \cite{2008ApJ...686.1383H} studied the two different mechanisms that have to be taken into account to explain the observation of prominences in various EUV coronal lines as dark features in the corona. The lowering of the brightness of EUV coronal lines at the prominence position can be due to two basic mechanisms: absorption and emissivity blocking (formerly called 'volume blocking'). The absorption of coronal line radiation at $\lambda < 912$~\AA\ is  due to the photoionisation of hydrogen, neutral and ionized helium. The lowering of coronal emission observed by Skylab was already interpreted  by absorption \citep{1976SoPh...50..365O}. 
\cite{2005ApJ...622..714A} derived theoretical expressions for the plasma opacity in different wavelengths leading to the determination of the column density of \ion{H}{i}, \ion{He}{i}, and \ion{He}{ii}.  The second mechanism is the emissivity blocking discussed by  \cite{2001ApJ...561L.223H} and theoretically developed by \cite{2005ApJ...622..714A}. This blocking which leads to a lower coronal emission is due to the presence of cool prominence plasma \citep{2004A&A...421..323S,2004SoPh..221..297S} or due to the low density of the hot plasma because of the existence of a cavity \citep{1994SoPh..149...51W,2010ApJ...724.1133G}. Simultaneous multiwavelength
images of a prominence at the limb show that the prominence can
appear in emission or in absorption depending on the wavelength of
the observations. Analysis of different EUV lines can help to
disentangle between these mechanisms  and, by inversion, to derive
physical quantities such as the opacity and column densities of
\ion{H}{i}, \ion{He}{i} and \ion{He}{ii} which constrain theoretical
models. In \cite{2008ApJ...686.1383H}, such analysis has been done
using multi wavelength observations during a coordinated campaign in
April 2007 (JOP 178\footnote{See \url{http://bass2000.obs-mip.fr/jop178/}.} and HOP 4\footnote{See \url{http://www.isas.jaxa.jp/home/solar/hinode_op/hop.php?hop=0004}.}).

In this paper we use spectra of the Hinode/EIS spectrometer
obtained during the same  campaign to show the
capability  of EIS to perform quantitative diagnostics of physical
parameters of prominences.
The EUV Imaging Spectrometer
embarked on the
Hinode mission \citep{2007SoPh..243....3K} provides us with the
possibility to observe the Sun in two wavelength bands
(170--211~\AA\ and 246--292~\AA). Each band has a distinct
1024$\times$2048 pixel CCD detector. We selected the prominence data
obtained on April 26, 2007.
The capabilities of the EIS spectrometer allow an
unprecedented view of the physical conditions within and around the
prominence, thanks to the large number of lines detected
across a range of temperatures. In far EUV, this is achieved for the first
time with spatial resolution around 1\arcsec.

The observations are detailed in Sect.~\ref{s:obs}. In the following sections, we present the rasters obtained  by integrating
the line profiles,  and the line profiles themselves. We discuss the
blends of the different lines and how they can be processed to
understand the images. The attenuation of the coronal emission in
the  presence of the prominence is interpreted by absorption and
emissivity blocking mechanisms.  In Sect.~\ref{s:he2-line} we
propose a method to derive the pure emission of the \ion{He}{ii}
256~\AA\ line which is blended by several coronal lines. 
The \ion{He}{ii} integrated  intensity is estimated under different assumptions (different scenarios for the interplay between the coronal blends and the absorption and blocking of the coronal radiation by the prominence plasma). The values obtained are then compared with theoretical values from 1D non-LTE modelling.

\section{Observations}\label{s:obs}

\begin{figure}
\centering
\resizebox{\hsize}{!}{\includegraphics{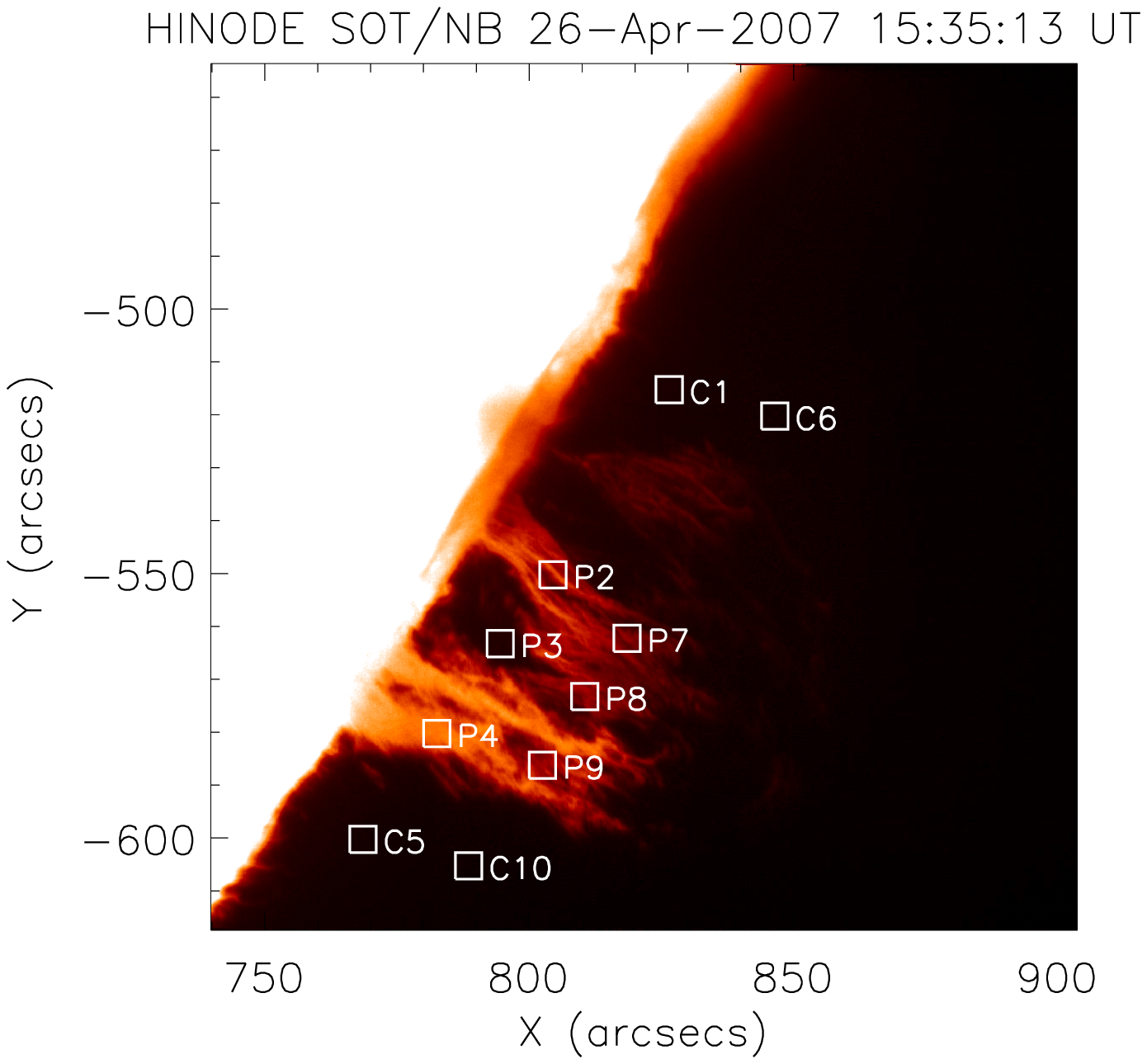}}
\resizebox{\hsize}{!}{\includegraphics{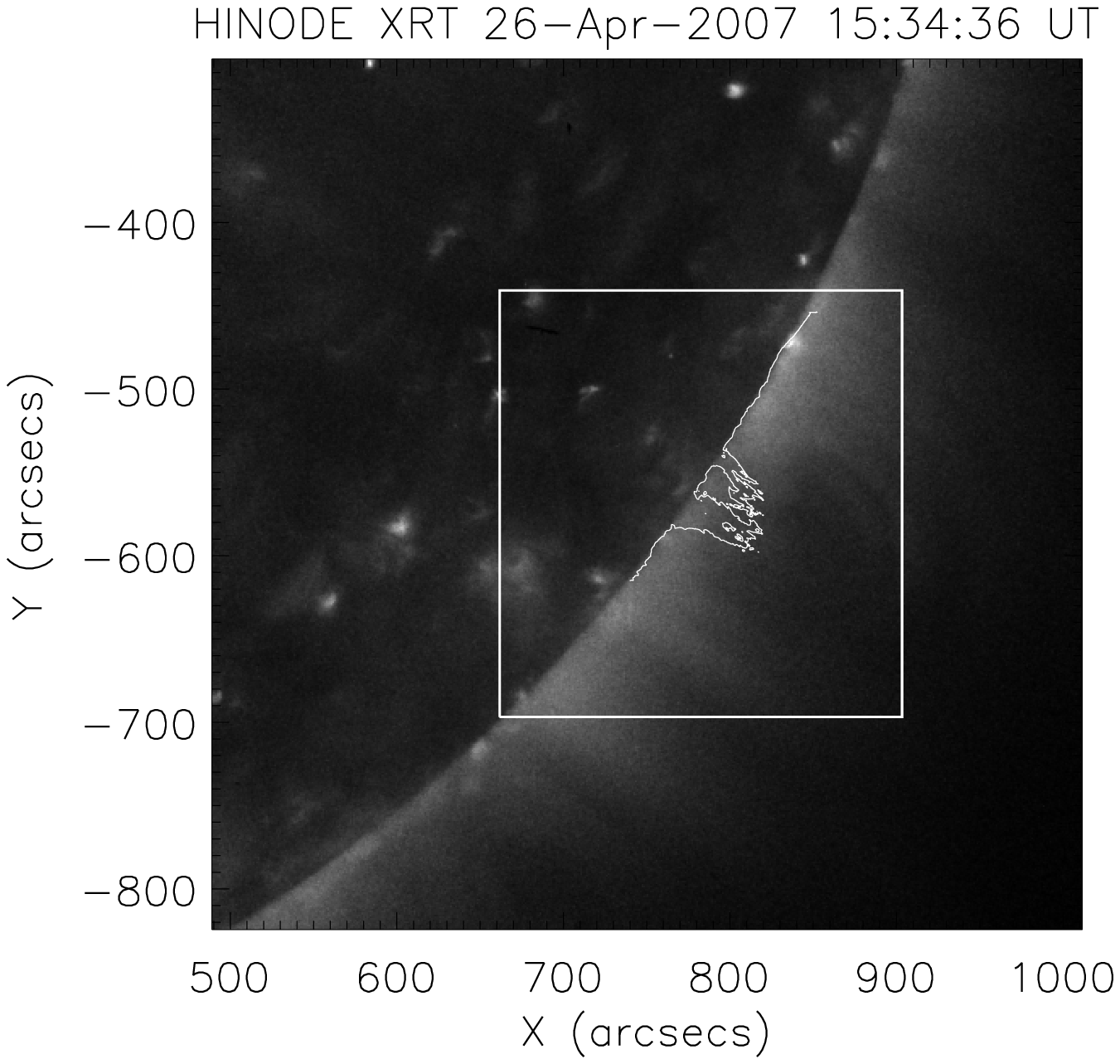}}
 \caption{(top) H$\alpha$ observation from Hinode/SOT of the west-limb prominence on 26 April
2007 at 15:35UT . The boxes on the SOT image indicate the positions around the prominence where the EIS line profiles are studied.
(bottom) Hinode/XRT image showing the prominence cavity. A contour of the H$\alpha$ limb including the prominence observed {with SOT} indicates the location of the prominence at the bottom of the cavity. {The white box corresponds to the region observed by Hinode/EIS.}}
 \label{sot}
\end{figure}

A  prominence surrounded by a large cavity was observed for three days (24-26 April 2007) during the campaign with Hinode (Fig.~\ref{sot}).
{This was possible because the corresponding filament extended  for  more than 30 degrees. During that time, different parts of the filament (footpoints, spine) crossed the disk and were observed as a prominence.}
On April 24, 2007, the prominence was already visible on the West limb in  the South  hemisphere between 30 and 35 degrees. The prominence  did not erupt even during the magnetic flux emergence at the basis of its cavity \citep{2009ApJ...704..485T}. The {long filament} was stable for two more days. \cite{2008ApJ...686.1383H} studied {a part of this long filament as a} prominence on April 25, 2007.
{The dynamics of this prominence over the period of observations was studied by \cite{2008ApJ...676L..89B,2010ApJ...716.1288B} who investigated the vertical flows observed by SOT, and also by \cite{2010A&A...514A..68S} who found evidence for flows inclined by an angle of 30 to 90 degrees in the prominence from \Ha\ observations.}

We focus here on the {last part of the filament. It consists of} well-formed off-limb structures seen on April 26, 2007, and is the best observed prominence in the EIS data set {for that sequence of observations}.
Here we analyze data from the \texttt{fil\_rast\_s2} study, which uses the 2\arcsec\ slit and makes a raster of 240\arcsec$\times$256\arcsec. At each position of the slit, a spectrum is obtained with a 50~s exposure. Data from 11 spectral windows is downloaded from the satellite: 10 narrow spectral windows, along with a full spectrum on the CCD~B of the EIS detector in the long wavelength region, between 189~\AA\ and 211~\AA\, (here referred to as CCDBLONG). This study enables us to observe chromospheric, transition region, and coronal features in a range of temperatures.
The selected spectral windows are indicated in Table~\ref{tab:eislines}.
The formation temperatures of the lines in this table are from \citet{1998A&AS..133..403M} and \cite{2007PASJ...59S.857Y}.
\begin{table} 
\caption{Lines observed in EIS rasters obtained with the \texttt{fil\_rast\_s2} study.}
        \label{tab:eislines}
        \centering
        \begin{tabular}{c c c l}
                \hline\hline
                Ion & Wavelength & $\log (T)$ & Blends\\
                \hline
                \ion{Fe}{viii} & 185.21 & 5.8 & \ion{Ni}{xvii}\\
                \ion{Fe}{viii} & 186.60 & 5.8 & \ion{Ca}{xiv}\\
                \ion{Ca}{xvii} & 192.82 & 6.7 & \ion{O}{v}, \ion{Fe}{xi}\\
                \ion{Fe}{xii}  & 195.12 & 6.1 & \ion{Fe}{xii}\\
                \ion{O}{v}     & 248.46 & 5.4 & \ion{Al}{viii}\\
                \ion{He}{ii}   & 256.32 & 4.7 & following four lines\\
                \ion{Si}{x} & 256.37 & 6.1& \\
                \ion{Fe}{x} & 256.40 & 6.0 & \\
                \ion{Fe}{xii} & 256.41 & 6.1 & \\
                \ion{Fe}{xiii} & 256.42 & 6.2 & \\
                \ion{Mg}{vi}   & 269.00 & 5.7 & \\
                \ion{Mg}{vi}   & 270.40 & 5.7 & \ion{Fe}{xiv}\\
      		    \ion{Fe}{xiv}  & 270.51 & 6.3 &\\
                \ion{Si}{vii}  & 275.35 & 5.8 & \\
                \ion{Mg}{v}\tablefootmark{a}    & 276.57 & 5.5 & \\
                \ion{Mg}{vii}  & 278.39 & 5.8 & \ion{Si}{vii}\\
                \ion{Mg}{vii}  & 280.75 & 5.8 &\\
                \hline
        \end{tabular}
        \tablefoot{The wavelengths (\AA), temperature of formation (K), {and known blends} are given.\\
        \tablefoottext{a}{Very weak line.}}
\end{table}

Using the provided software from SolarSoft, we corrected the data for several effects (dark current subtraction, cosmic ray removal, flat field correction, and flagging hot pixels), and convert the number of counts into physical intensity units. Additional corrections to the data consist in internal co-alignment of the two detectors, slit tilt, and orbital variation of the line centroids. We use the routines written by P.R.~Young available in SolarSoft to correct for these effects.

Figure~\ref{sot} shows this prominence observed by Hinode/SOT in the H$\alpha$ line and the surrounding cavity   observed by XRT. 
{The temporal evolution of the prominence during the EIS raster between 14:46UT and 16:29UT can be seen in the SOT movie associated with on-line Figure~\ref{f:movie}.}
\onlfig{2}{
\begin{figure}
\resizebox{\hsize}{!}{\includegraphics{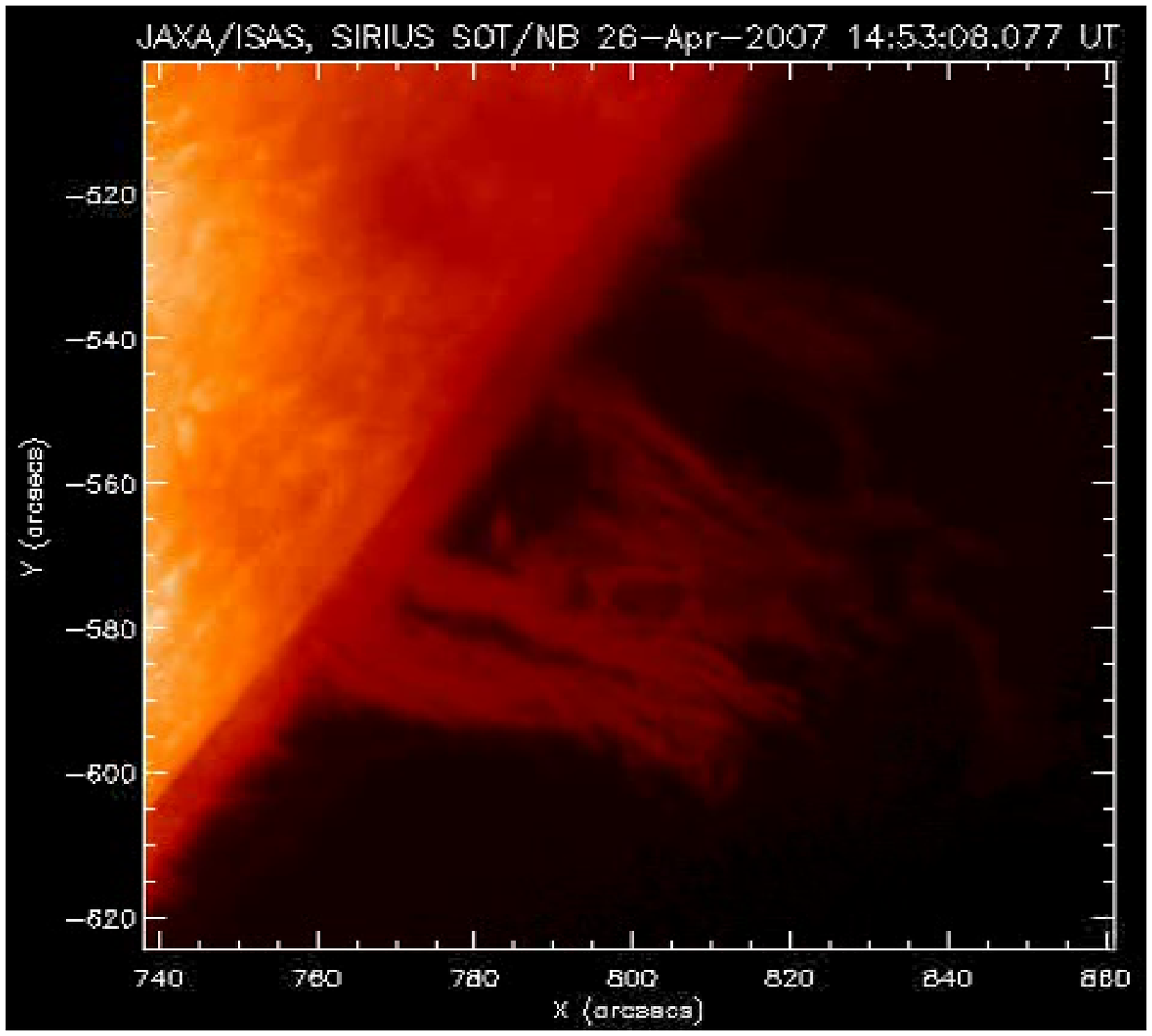}}
\caption{A still from the Hinode/SOT \Ha\ movie showing the temporal evolution of the prominence observed on 26 April 2007 during the EIS raster between 14:46UT and 16:29UT.} \label{f:movie}
\end{figure}
}
The prominence on that day is more active and changes shape more rapidly than on the previous days. {In the movie} we see rather thin and dynamical H$\alpha$ bright vertical structures separated by large dark  lanes and reaching a height of $\sim 40 000$~km (60\arcsec). We have selected 10 boxes ($5\times5\arcsec$) in {and around} the prominence at two values of the height above the solar limb as seen in \Ha\ ($H_1=20$~Mm and $H_2=38$~Mm),  corresponding to different structures of the prominence {and its environment}. We will study the behaviour of the intensities of the different lines observed by EIS in these 10 boxes (integrated intensity and profile) in the next sections.
The cavity is visible but fainter than on April 25 \citep{2008ApJ...686.1383H}. It extends to more than 100\arcsec over the limb.

\section{Integrated intensities and rasters}

 \subsection{EUV images}

\begin{figure}
    \centering
    \resizebox{\hsize}{!}{\includegraphics{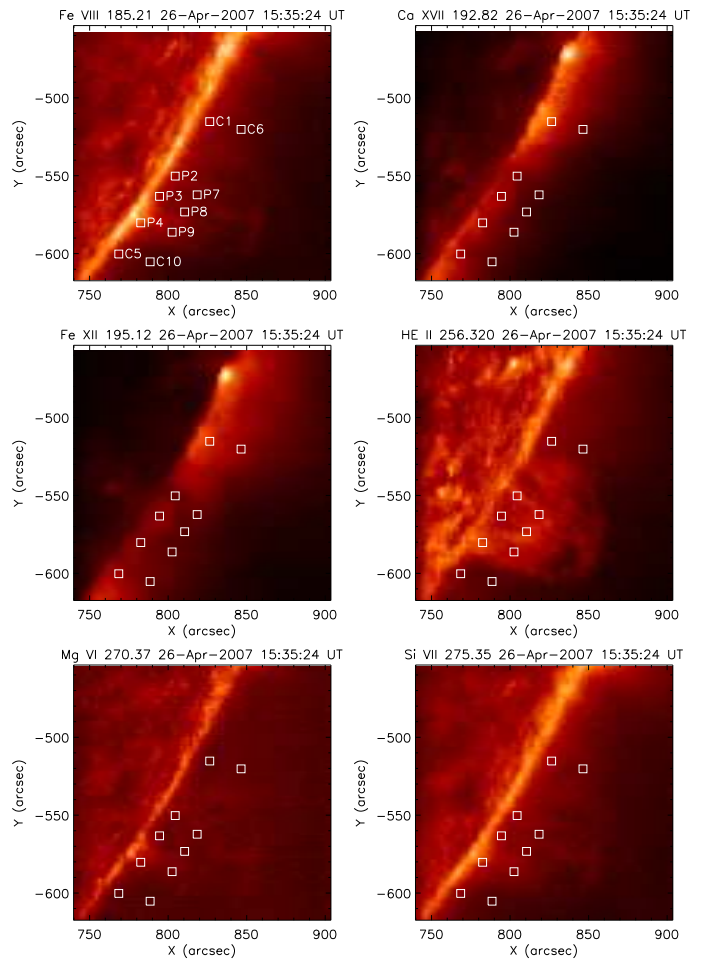}}
    \caption{Raster images obtained with \texttt{fil\_rast\_s2} on April 26, 2007
    between 14:46UT and 16:29UT. The white boxes on each raster indicate the
    positions around the prominence where line profiles are averaged.}
    \label{rasters}
\end{figure}

  Several interesting lines are present in the EIS observations
such as the \ion{Fe}{xii} 195.12~\AA\ line -- which is one of the
three core lines along with \ion{Ca}{xvii} 192.82~\AA\ and
\ion{He}{ii} 256.32~\AA\ that each EIS study must include. Some of
the raster images are shown in Fig.~\ref{rasters}. 
The prominence is
best identified in the \ion{He}{ii} 256.32~\AA\ line, which is the
coolest line in this data set. In other cool lines (\eg\ \ion{O}{v})
with $\log(T)=5.4-5.5$, the prominence is much less visible (not
shown here). This is explained by lower count rates on the detector.
The signatures of the prominence in the corona are different
according to the temperature of formation of the lines.
 There are
basically three categories of lines detected by EIS within the
prominence. In coronal lines the prominence appears to be darker
than the surrounding corona, while in TR lines it is brighter than the corona,
showing namely the prominence-corona transition region (PCTR).
Finally, there is the \ion{He}{ii} 256~\AA\ resonance line, which is bright on
the disk chromosphere and transition region, and is well detected also within the
prominence.

On each raster we show several boxes corresponding to different regions where the line profiles (presented in the following sections) are averaged. The boxes are identified by the letter C when the box is located in the corona outside the prominence, and {by the letter} P when the box is on the prominence.
P-3 is located in a 'hole' of the prominence (a region of the corona almost empty of prominence material) as seen in H$\alpha$ (see Fig.~\ref{sot}). This kind of void under the prominence is evolving fast {(see on-line Figure~\ref{f:movie} and associated movie)}.

The analysis of the rasters combined with the line profiles allows
us to obtain a good qualitative as well as quantitative diagnostics.
A careful inspection of these images reveals several
interesting features.

The \ion{He}{ii} prominence  plasma is in a loop-like
shape and reaches higher heights than  the H$\alpha$ vertical
structures. Some material can be seen in \ion{He}{ii} and not in
H$\alpha$. Is this due to a difference of temperature or density? Do we observe resonance scattering or PCTR emission? Non-LTE models can help us to address  these questions. Due to the blend of this line by several
coronal lines, a careful reconstruction (deblending) is needed before any further
analysis (see Sect.~\ref{s:he2-line}).

{In this data set,} the observed lines are formed within a large temperature range, from cool TR (\ion{He}{ii} at $\log T=4.7$) to coronal temperatures (\ion{Fe}{xiv} at $\log T=6.3$) -- see Table~\ref{tab:eislines}. The prominence has a cool core (typically 6000 to 8000~K) surrounded by a PCTR. The core of the prominence can be seen in emission in H$\alpha$, while EUV lines are emitted in the PCTR. Coronal line emission in this wavelength range can be absorbed due to the photoionisation of \ion{He}{i} for $\lambda <  504$~\AA\ (\ion{He}{i} continuum head) and due to the photoionisation of \ion{He}{i} and \ion{He}{ii} for $\lambda<  228$~\AA\ (\ion{He}{ii} continuum head). Over the range of EIS wavelengths, the photoionization of hydrogen is less important \citep[see][for all relevant cross-sections]{2005ApJ...622..714A}. The lack of coronal emission can also be due to the presence of the cavity (emissivity blocking).

\subsection{Integrated intensities}

In order to quantify the lowering of coronal lines and the emission of the PCTR of the prominences, integrated intensities have been computed in the 10 boxes. Table~\ref{tab:intensities} presents the integrated intensities (in erg s$^{-1}$ cm$^{-2}$ sr$^{-1}$) {for several lines observed by EIS} in various parts of the prominence and its surroundings at {two} values of the height above the solar limb: $H_1=20$~Mm, $H_2=38$~Mm. 
\begin{table*}
    \caption{Integrated line intensities in various parts of the prominence and its surroundings.}
        \label{tab:intensities}
        \centering
        \begin{tabular}{l|ccccc|ccccc}
                \hline\hline
                Line & \multicolumn{5}{c}{$H_1=20$~Mm} & \multicolumn{5}{c}{$H_2=38$~Mm}\\
                & C-1 & P- 2 & P- 3 & P-4 & C-5 & C-6 & P-7 & P-8 & P-9 & C-10\\
                \hline
     \ion{Fe}{viii} 185.21 & 71 & 76 & 84 & 136 & 58 & 35 & 47 & 47 & 56 & 24 \\
           & (101) & (103) & (109) & (161) & (85) & (60) & (66) & (73) & (78) & (45) \\
     \ion{Fe}{viii} 186.60 & 56 & 61 & 60 & 103 & 44 & 28 & 34 & 37 & 45 & 19 \\
           & (73) & (76) & (78) & (121) & (59) & (43) & (47) & (51) & (57) & (31) \\
     \ion{Ca}{xvii} 192.82 & 216 & 143 & 179 & 190 & 175 & 119 & 95 & 100 & 114 & 102 \\
           & (228) & (151) & (187) & (201) & (181) & (125) & (99) & (103) & (118) & (105) \\
     \ion{Fe}{xii} 195.12 & 927 & 588 & 723 & 691 & 787 & 681 & 520 & 482 & 534 & 575 \\
           & (959) & (604) & (742) & (716) & (805) & (701) & (532) & (495) & (546) & (582) \\
     \ion{O}{v} 248.46 & 44 & 33 & 34 & 46 & 29 & 28 & 27 & 33 & 30 & 20 \\
           & (103) & (99) & (98) & (128) & (104) & (95) & (94) & (96) & (96) & (102) \\
     \ion{He}{ii} 256.32 & 360 & 383 & 355 & 489 & 316 & 224 & 323 & 353 & 328 & 191 \\
           & (404) & (423) & (404) & (538) & (355) & (264) & (360) & (392) & (368) & (231) \\
     \ion{Mg}{vi} 270.37 & 39 & 26 & 32 & 34 & 34 & 44 & 32 & 28 & 27 & 30 \\
           & (55) & (42) & (48) & (52) & (50) & (60) & (47) & (43) & (44) & (46) \\
     \ion{Si}{vii} 275.35 & 43 & 46 & 49 & 79 & 36 & 21 & 26 & 31 & 32 & 12 \\
           & (59) & (62) & (67) & (94) & (51) & (37) & (40) & (44) & (49) & (29) \\
     \ion{Mg}{vii} 278.39 & 37 & 40 & 42 & 76 & 31 & 21 & 22 & 27 & 29 & 10 \\
           & (57) & (60) & (63) & (96) & (51) & (39) & (41) & (46) & (47) & (29) \\
     \hline
     XRT & 19 & 14 & 14 & 15 & 16 & 13 & 11 & 10 & 11 & 11 \\
     \hline
        \end{tabular}
        \tablefoot{Intensities are in erg s$^{-1}$ cm$^{-2}$ sr$^{-1}$ for EIS, and in arbitrary units for XRT. C- and P- correspond  respectively to boxes in the corona and in the prominence. P-3 corresponds to the hole seen in H$\alpha$. {For each EIS line, the first row corresponds to intensity values with a constant background removed from the spectrum. The second row (values between brackets) has no background subtraction.} The uncertainty in radiometric calibration {of EIS data} is of the order of 22\% \citep{2006ApOpt..45.8689L}.}
\end{table*}
For each {EIS} line, the first row corresponds to intensity values {where a constant background was removed from the spectrum.}
This background is supposed to be of instrumental origin, as it is roughly constant with position for any particular spectral line.
The second row has  no background subtraction.  
{The XRT intensities (in arbitrary units) are also indicated in Table~\ref{tab:intensities}. The full variation of the XRT intensity at the two heights $H_1$ and $H_2$ is also shown in Fig.~\ref{cuts}.}

Table~\ref{tab:percent} presents  the ratios between the intensities at the prominence position to that outside in the corona {for several EUV TR and coronal emission lines, as well as for the XRT intensities. This allows us to calculate the optical thickness at 195~\AA\ in the same way as in \cite{2008ApJ...686.1383H} (see Sect.~\ref{s:abs}).}
\begin{table*}
    \caption{Ratios between the  intensities of transition region and coronal lines measured on the prominence and in the corona.}
    \label{tab:percent}
    \centering
    \begin{tabular}{l c c c c}
    \hline\hline
    Line  & \multicolumn{2}{c}{$H_1$} & \multicolumn{2}{c}{$H_2$}\\
    &   $I(P2)/I(C1)$ & $I(P4)/I(C5)$ & $I(P7)/I(C6)$ & $I(P9)/I(C10)$\\
    \hline
    & \multicolumn{4}{c}{Increase of TR emission}\\
    \ion{Fe}{viii} 185.21 & 1.07 & 2.34 & 1.34 & 2.33\\
    \ion{Fe}{viii} 186.60 & 1.09 & 2.34 & 1.21 & 2.37\\
    \ion{Si}{vii}  275.35 & 1.07 & 2.19 & 1.24 & 2.67\\
    \hline
    & \multicolumn{4}{c}{Decrease of coronal emission}\\
    \ion{Fe}{xii} 195.12 ($r$)	& 0.63 & 0.88 & 0.76 & 0.93\\
    XRT ($r_b$)			& 0.74 & 0.94 & 0.85 & 1.00\\
    \hline
    $\tau_{195}$ ($\alpha=0.5$) & 0.33 & 0.14 & 0.22 & 0.15\\
    $\tau_{195}$ ($\alpha=0.3$) & 0.62 & 0.24 & 0.39 & 0.27\\
    \hline
    \end{tabular}
    \tablefoot{$r_b$ is the ratio between coronal emission at a position in the prominence to a position outside in XRT data. The optical thickness at 195~\AA\ is derived by assuming a spherically symmetric corona with the prominence lying in the plane of the sky ($\alpha=0.5$) and for an asymmetric corona with the prominence lying behind the plane of the sky ($\alpha=0.3$).}
\end{table*}
The intensity of the TR lines \citep[comparable to those observed by][]{2010ApJ...714..636L} is enhanced by a factor of 1--2 due to the emission from the prominence-corona transition region.
The decrease of the intensity of coronal lines is due to the presence of the prominence {and of the surrounding cavity}, and is therefore more significant at low heights where the prominence is denser. Note also that P-4 is in a bright region above the limb.

\section{Coronal lines}

\begin{figure}
    \centering
    \resizebox{\hsize}{!}{\includegraphics{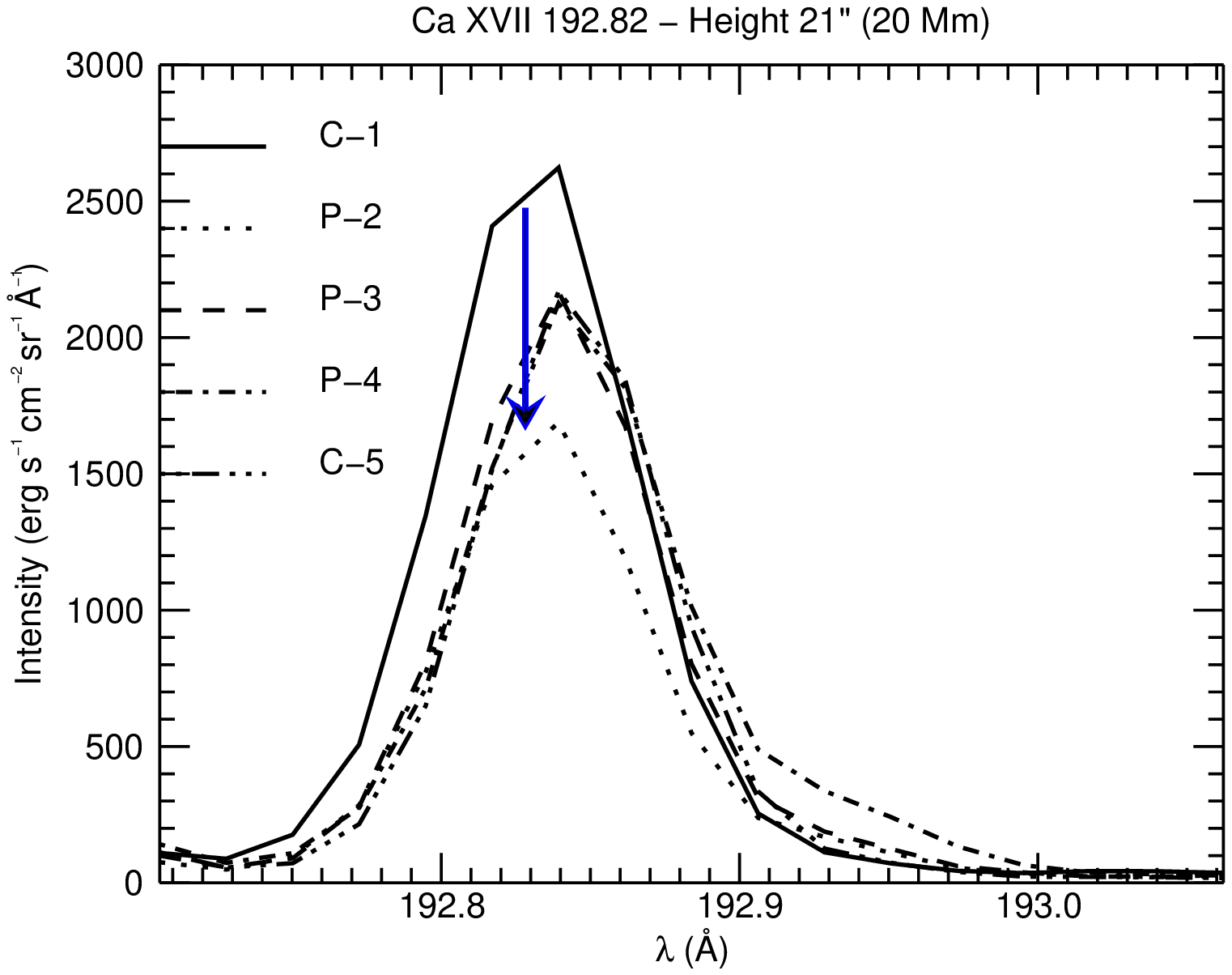}}
    \resizebox{\hsize}{!}{\includegraphics{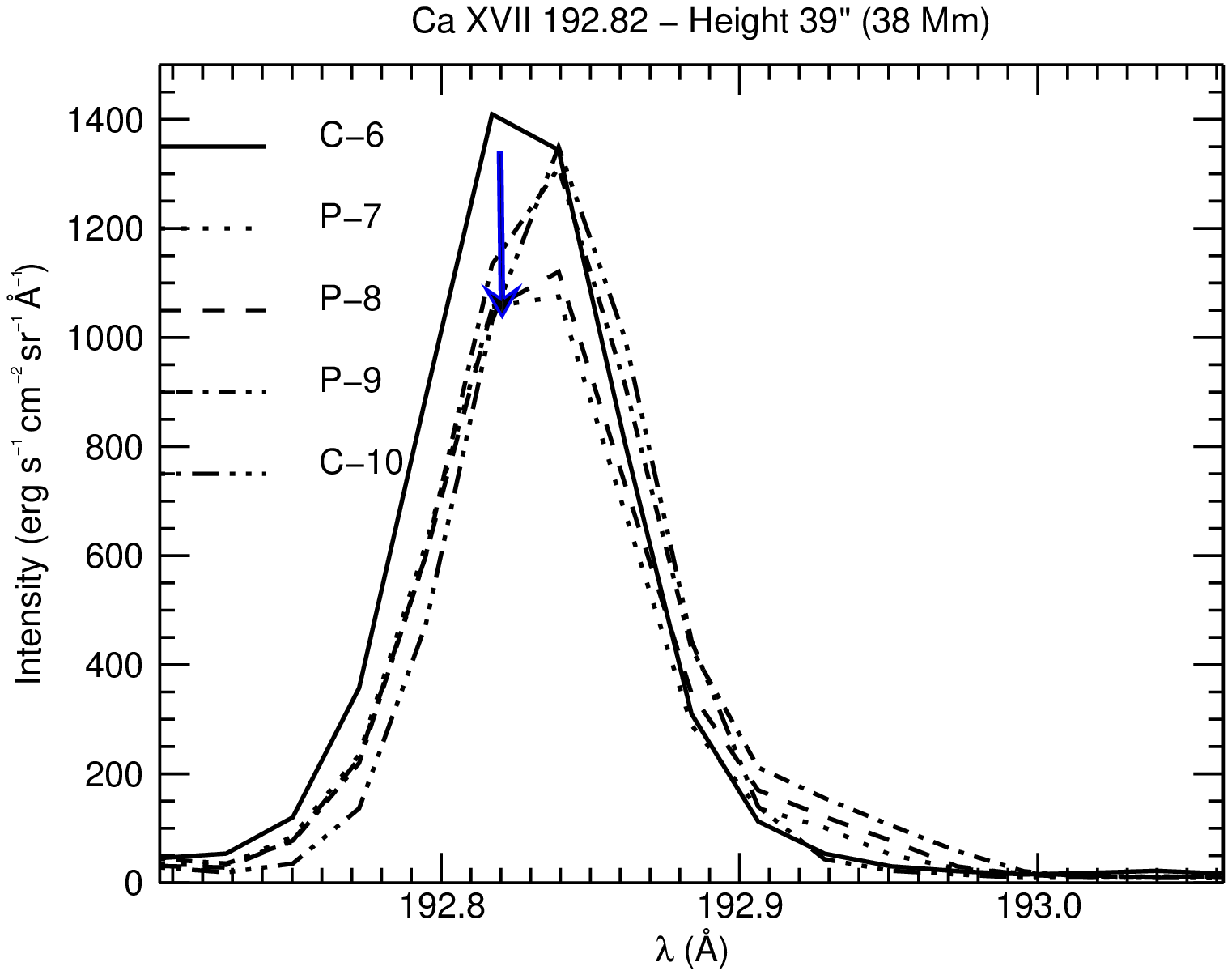}}
    \caption{\ion{Ca}{xvii} 192 line profiles obtained with \texttt{fil\_rast\_s2} on
    April 26, 2007 between 14:46UT and 16:29UT at two heights (20000~km and 38000~km).
    The downward arrows indicate the decrease in line emission from the corona to the prominence.
    Boxes 1, 5, 6, 10 are in the corona, while the other ones are located on the prominence.}
    \label{prof192}
\end{figure}

\begin{figure}
    \centering
    \resizebox{\hsize}{!}{\includegraphics{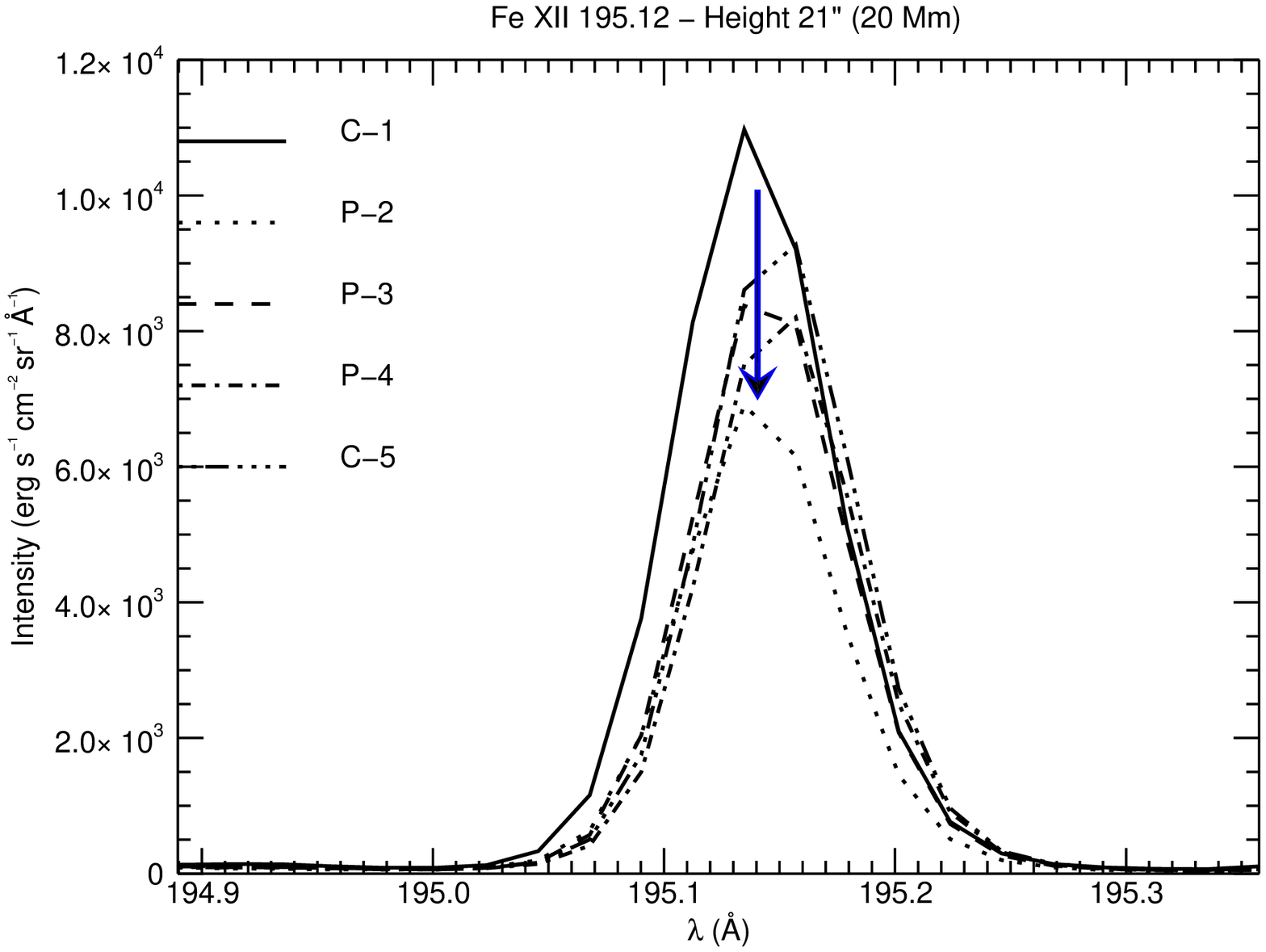}}
    \resizebox{\hsize}{!}{\includegraphics{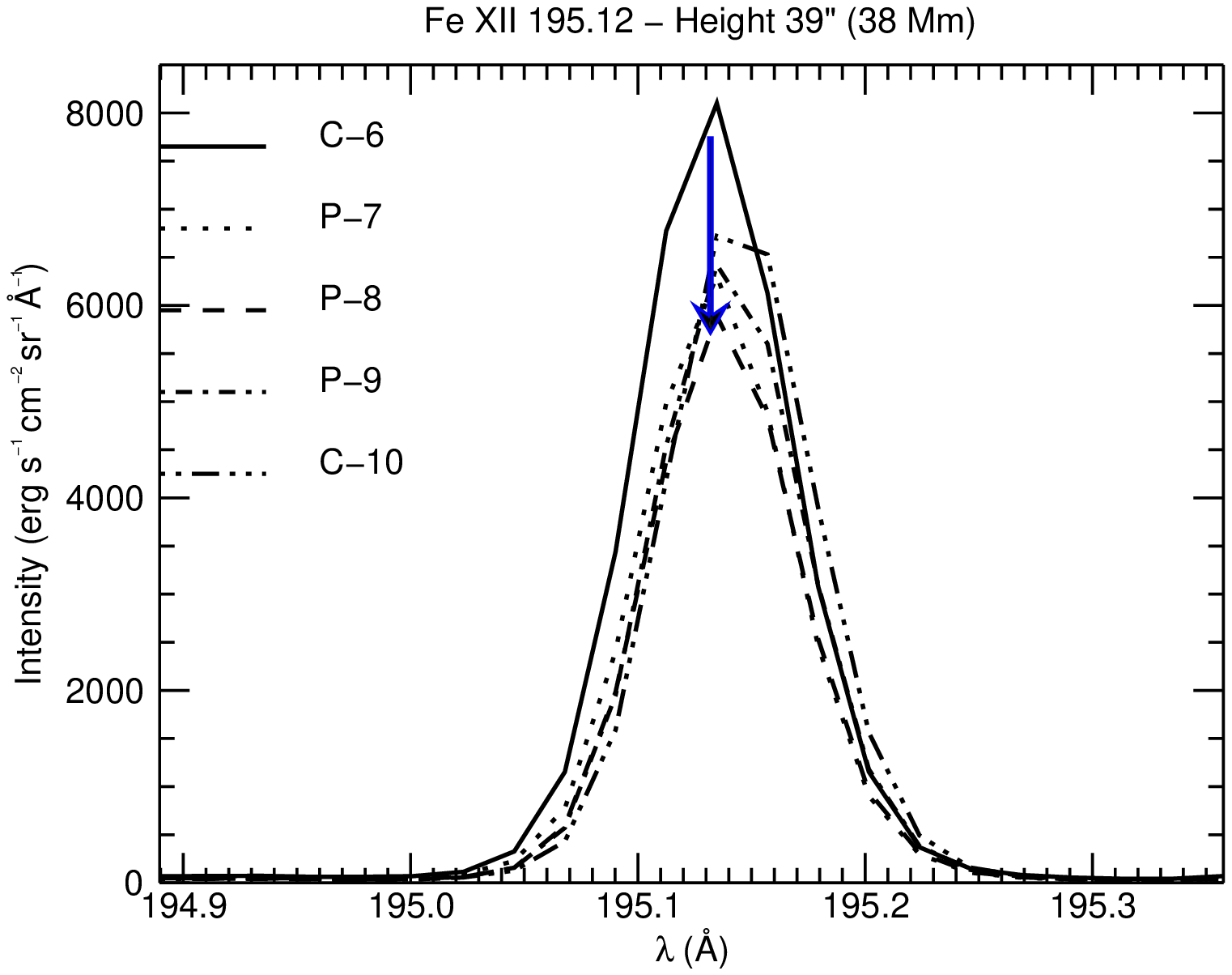}}
    \caption{{Same as Fig.~\ref{prof192} for} \ion{Fe}{xii} 195 line profiles.
    The downward arrows indicate the decrease in line emission from the corona to the prominence.}
    \label{prof195}
\end{figure}

On each panel of Figs.~\ref{prof192} and \ref{prof195}, we  plot the
profiles of the \ion{Ca}{xvii} 192.82~\AA\ and \ion{Fe}{xii} 195.12~\AA\ lines for several boxes corresponding to different locations
on the prominence (P-2, P-3, P-4, P-7, P-8, P-9) or in the corona (C-1, C-5, C-6, C-10).

\subsection{Emissivity blocking}\label{s:blocking}

At XRT wavelengths, the absorption by resonance continua is
negligible \citep{2007SoPh..242...43A} and the low emission in the
corona is entirely due to the presence of the cavity and the
prominence itself. The cavity has a low density plasma and the
prominence cool plasma blocks a certain volume. A mechanism
that can explain the dark void is  the emissivity
blocking mechanism \citep{2008ApJ...686.1383H}.

\begin{figure}
\centering 
\resizebox{\hsize}{!}{\includegraphics{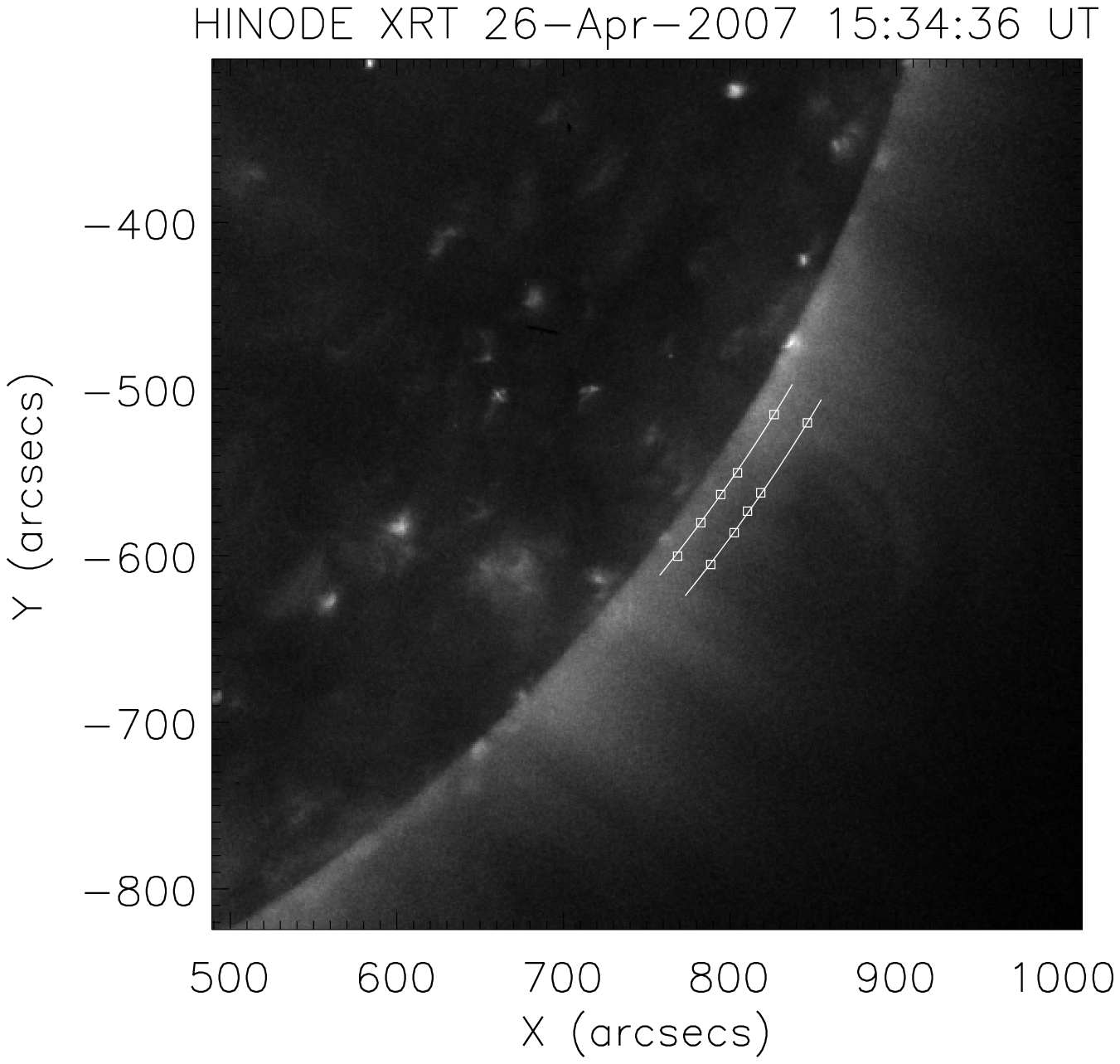}}
\resizebox{\hsize}{!}{\includegraphics{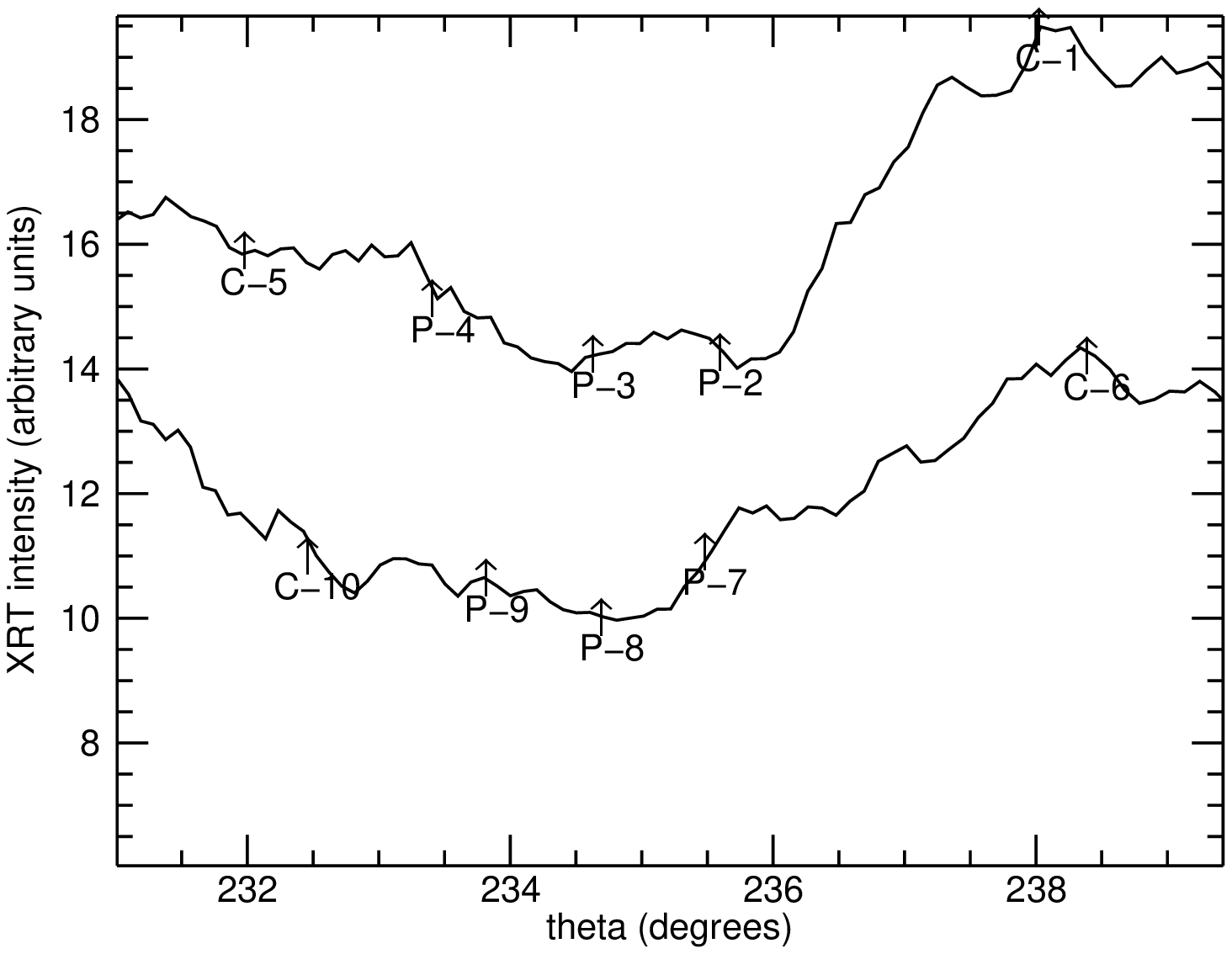}}
\caption{XRT image {(top)} and cuts {(bottom)} through the cavity.}
 \label{cuts}
\end{figure}

In order to estimate the level of emissivity blocking, we proceed in the same way as in \cite{2008ApJ...686.1383H}. In the top panel of Fig.~\ref{cuts} we show {two} cuts in the XRT image, made {at constant} heights and {over about 8 degrees}.
 The relative intensities for these {two} cuts are
displayed in the bottom panel of Fig.~\ref{cuts}. We may compare the brightness in the
darkest part of the cavity with that at cavity boundaries (i.e.
'quiet' corona). As a result, we get the emissivity-blocking
factors {$r_b$ shown in Table~\ref{tab:percent}}.
The cavity is most pronounced at larger altitudes as
evident from Fig. \ref{cuts}. 

\subsection{Absorption}\label{s:abs}

The \ion{Fe}{xii} line at 195.12~\AA\ has a strong integrated intensity between 495 and 959 erg s$^{-1}$ cm$^{-2}$ sr$^{-1}$ {(background included)}. This line is broadened due to a blend by another line of \ion{Fe}{xii} at 195.18 \AA\ which has the effect to increase the intensity.  The absorption mechanism is efficient at this wavelength and it partially explains the dark features observed  in the \ion{Fe}{xii} 195.12 raster (Fig.~\ref{rasters}). Fig.~\ref{prof195} presents the line profiles. The intensity in  boxes C-1 and C-5 corresponding to the corona (I$_B$) is higher than in the prominence boxes P-2 -- P-4 (I$_P$).  The attenuation of the intensity is due to the partial absorption of the coronal radiation behind the prominence and to the emissivity blocking which should be taken into account. 
In the general case, the opacity of the prominence can be estimated by the following formula \citep{2008ApJ...686.1383H}:
\begin{equation}
\tau = - \ln \left( \frac{r}{\alpha r_b} - \beta \right) \ ,
\end{equation}
{where $r=I_P/I_B$ is the ratio of the intensity at the prominence position to that outside at 195~\AA, and $r_b$  is the ratio between coronal emission at a position in the prominence to that outside in XRT data (where negligible absorption takes place). 
The background radiation intensity $I_B$ at the prominence position is that derived from XRT data multiplied by a factor $\alpha$, while the foreground intensity will be a fraction $(1-\alpha)$ of the XRT intensity, and $\beta=(1-\alpha)/\alpha$.
Assuming that the coronal line emissivity is symmetrically distributed behind and in front of the prominence, then $\alpha=0.5$.
We estimate that the bulk of the prominence is situated behind the plane of the sky, as it had been crossing the limb two days before these observations. Therefore we also estimate the optical depth at 195~\AA\ for the case of an asymmetric corona with $\alpha=0.3$.
The optical depth at 195~\AA\ inferred in this way {and shown in Table~\ref{tab:percent}} is in good agreement with the values obtained by \cite{2008ApJ...686.1383H}. 
Note that these results do not depend on the position of our coronal box. As long as the same box is chosen for the reference intensities in XRT and at 195~\AA, the inferred optical depth at 195~\AA\ in the prominence box will be the same.}

{The procedure adopted here to derive the optical thickness at 195~\AA\ is similar to the one used later in this work in Sect.~\ref{s:deblend}, with the difference that here we disentangle between absorption and emissivity blocking.}

The intensity of the \ion{Fe}{xii} line in box P-3 (\Ha\ void) is larger than  in P-2 and P-4, but is lower than in C-1 and C-5. Is it due to absorption by optically thin prominence plasma not well visible in H$\alpha$? The answer is not clear, even with the information from the intensities of transition region lines such as the \ion{Si}{vii} 275 line.

\section{Transition region lines}

\subsection{Fe and Si lines}

\begin{figure}
    \centering
    \resizebox{\hsize}{!}{\includegraphics{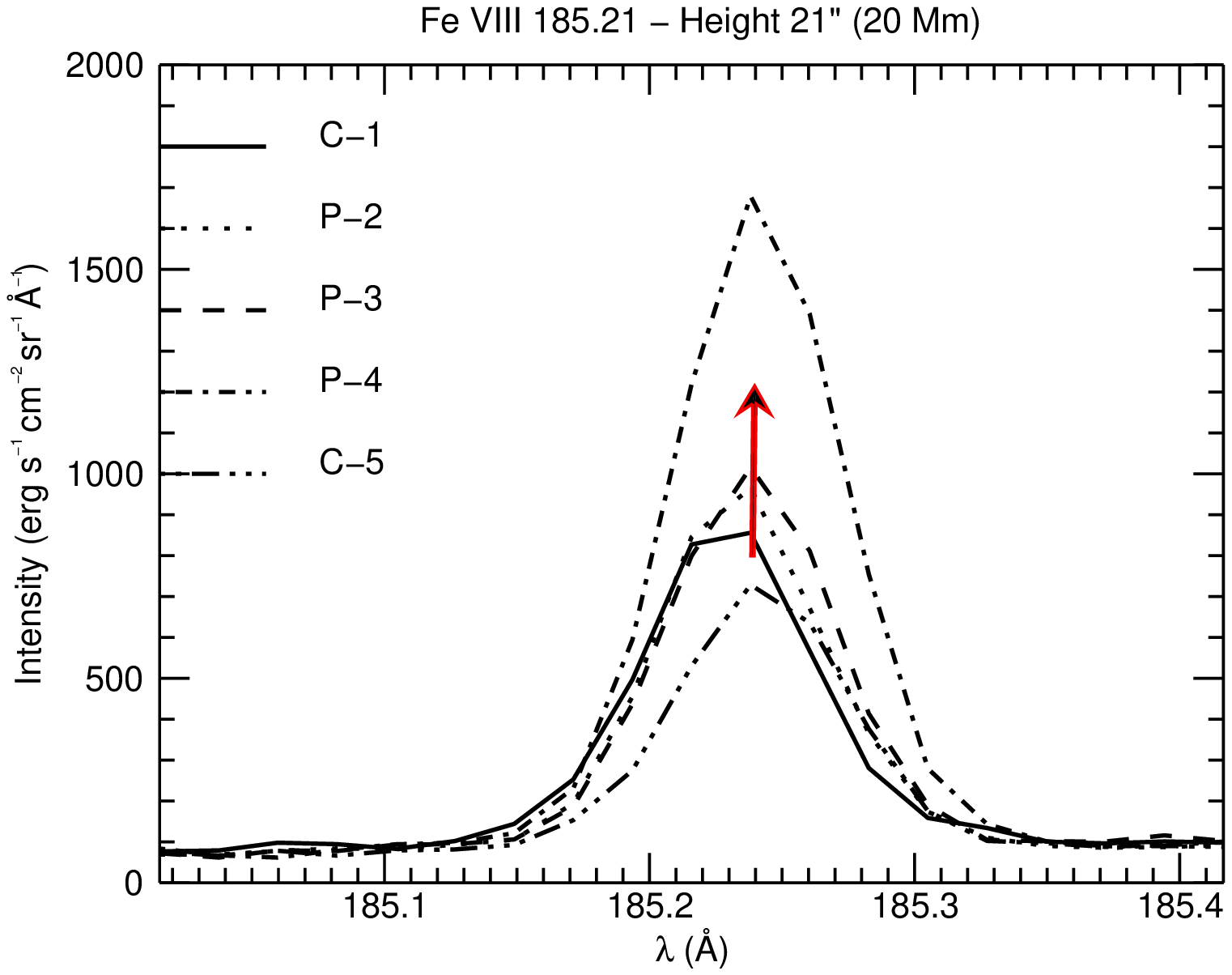}}
    \resizebox{\hsize}{!}{\includegraphics{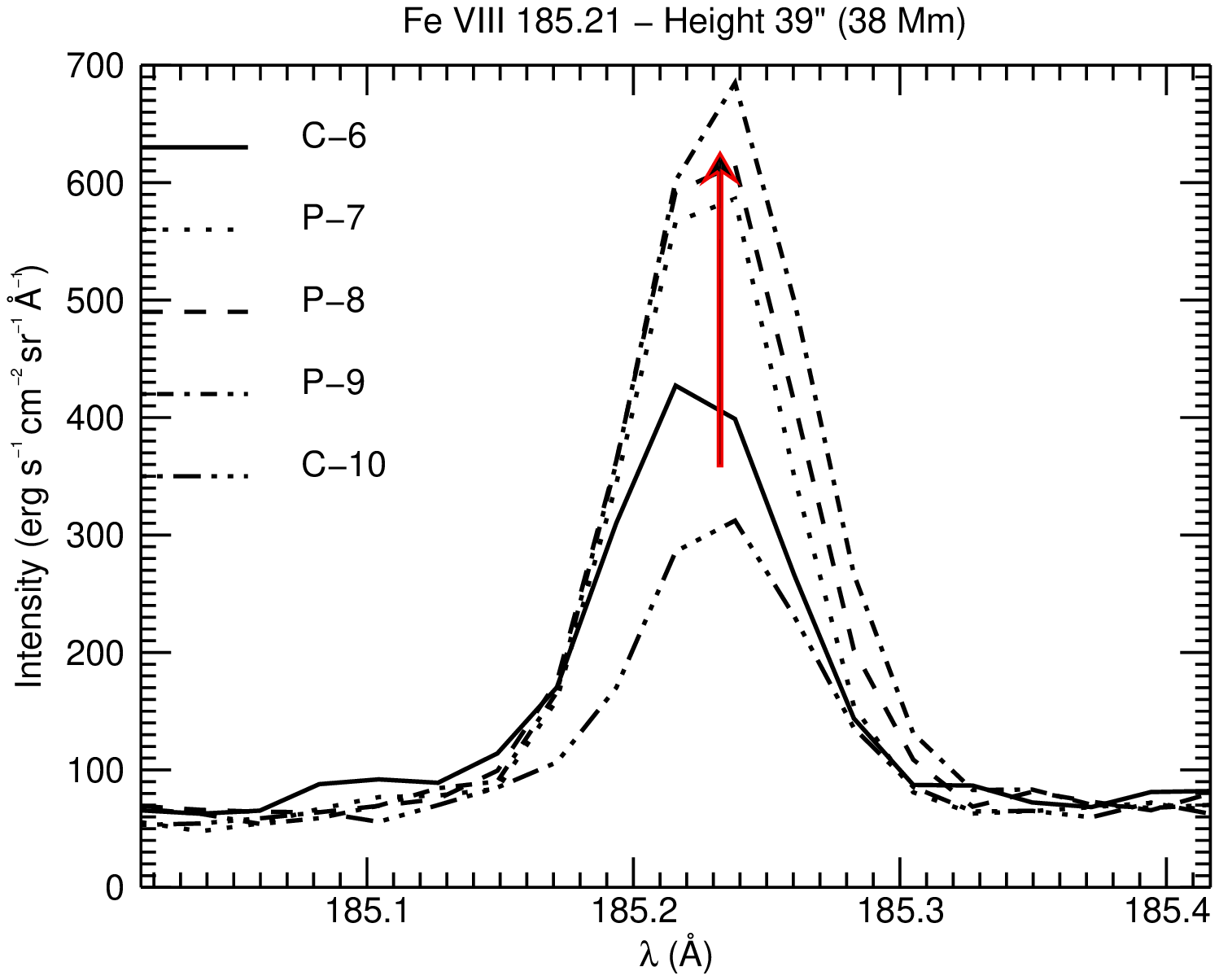}}
    \caption{{Same as Fig.~\ref{prof192} for} \ion{Fe}{viii} 185 line profiles.
    The upward arrows indicate the increase in line emission from the corona to the prominence.}
    \label{prof185}
\end{figure}

\begin{figure}
    \centering
    \resizebox{\hsize}{!}{\includegraphics{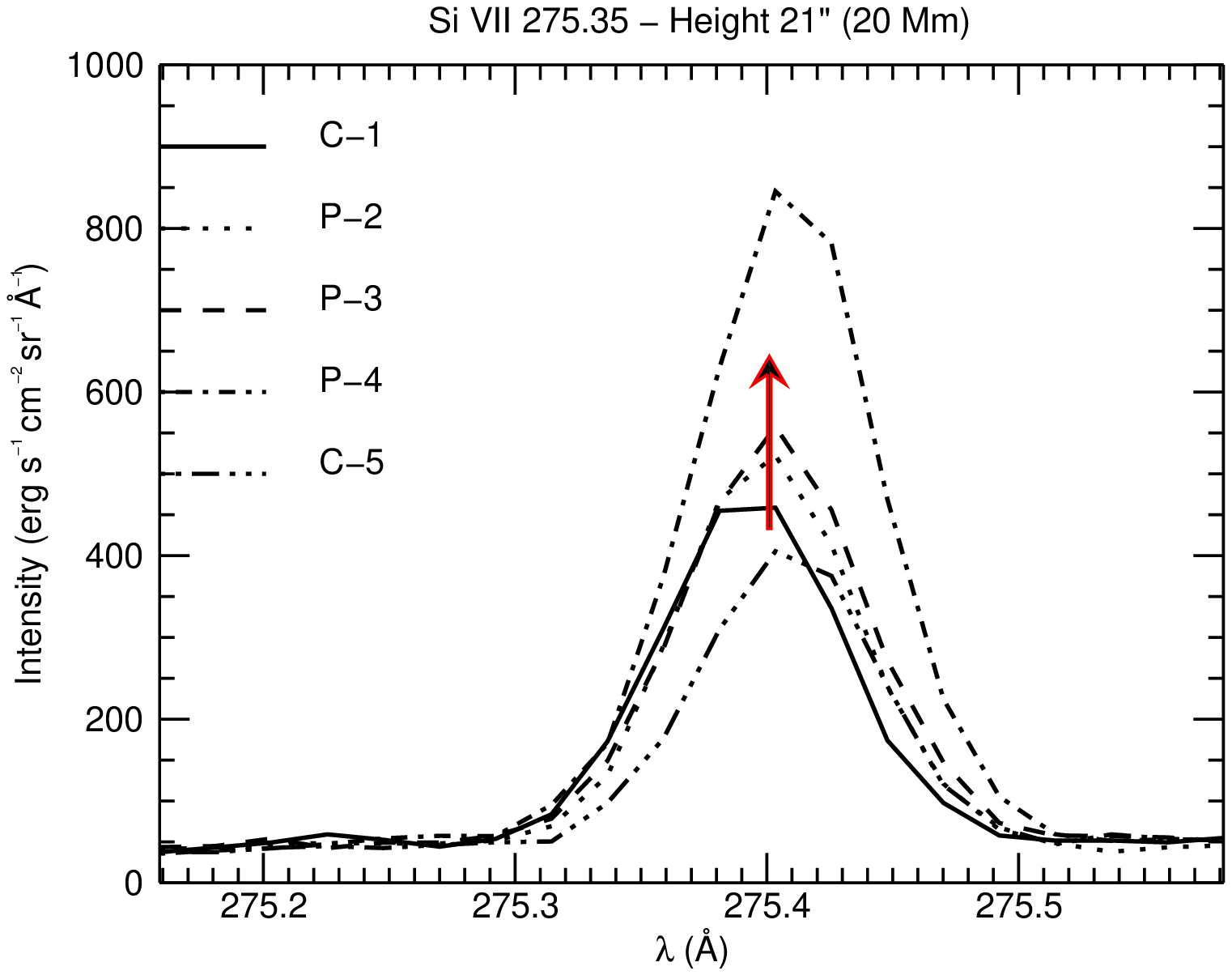}}
    \resizebox{\hsize}{!}{\includegraphics{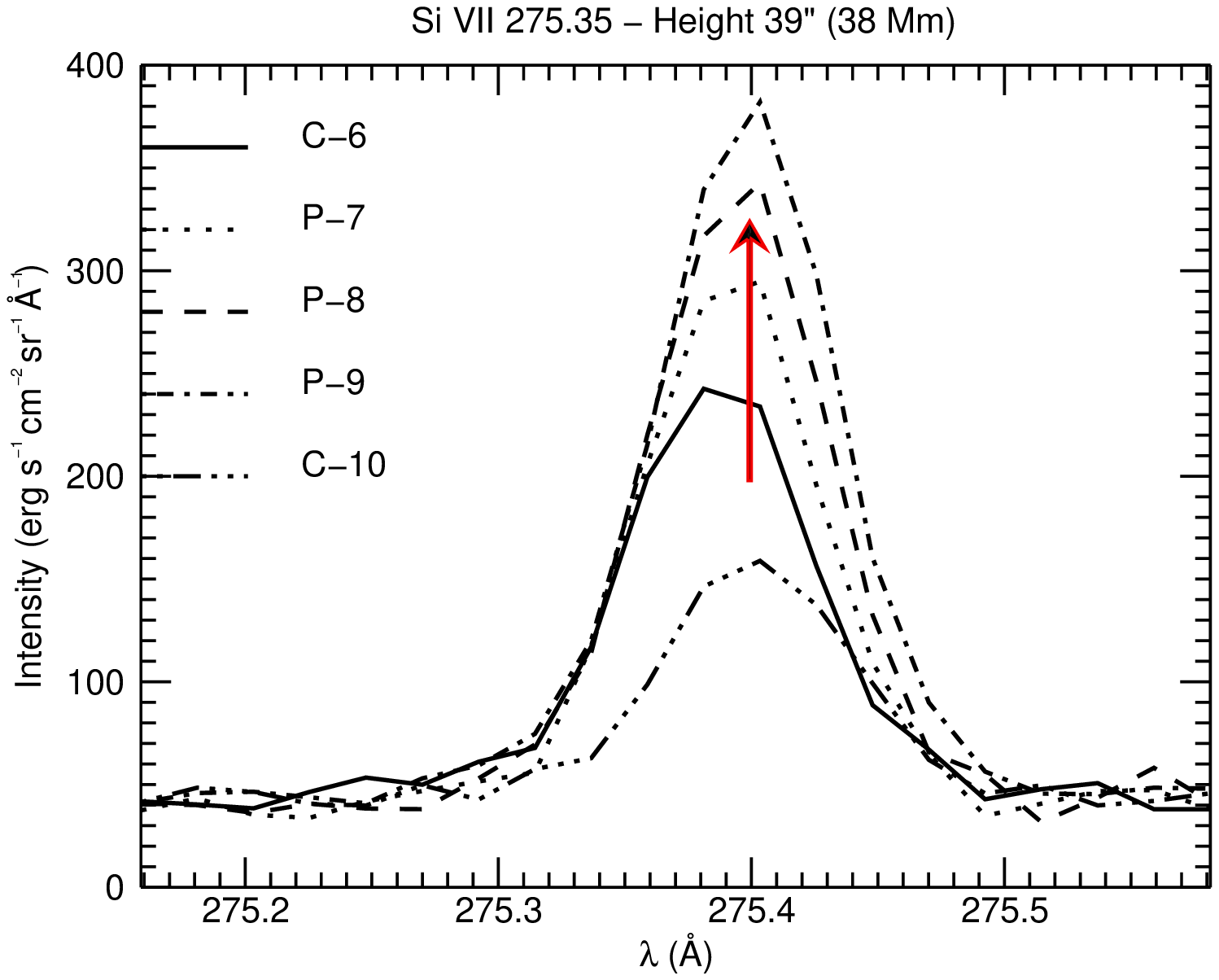}}
    \caption{{Same as Fig.~\ref{prof192} for} \ion{Si}{vii} 275 line profiles.
    The upward arrows indicate the increase in line emission from the corona to the prominence.}
    \label{prof275}
\end{figure}

The prominence  is also clearly identified in emission against the
background corona in the \ion{Fe}{viii} 185.21~\AA\ and \ion{Si}{vii} 275.35~\AA\  lines
(Fig.~\ref{rasters}) formed at $\log(T)\sim5.8$ \citep{2007PASJ...59S.727Y}. It is also visible in  
\ion{Fe}{viii} 186.60~\AA\ (not shown). 
Corresponding line profiles are shown in Figs.~\ref{prof185} and \ref{prof275}.
The  \ion{Si}{vii} line has no known blend. Note that the {shape and intensity} of the line profile in box C-10 {are} very different from {those in} box C-6.

In these  lines we are detecting the prominence-to-corona transition region. {Although} the \ion{Fe}{viii} lines are blended by coronal lines (see Table~\ref{tab:eislines}), the intensity of these coronal lines is low enough that we see mainly the emission of \ion{Fe}{viii}  lines coming from the PCTR.

\subsection{\ion{Mg}{vi} line}

\begin{figure}
    \centering
    \resizebox{\hsize}{!}{\includegraphics{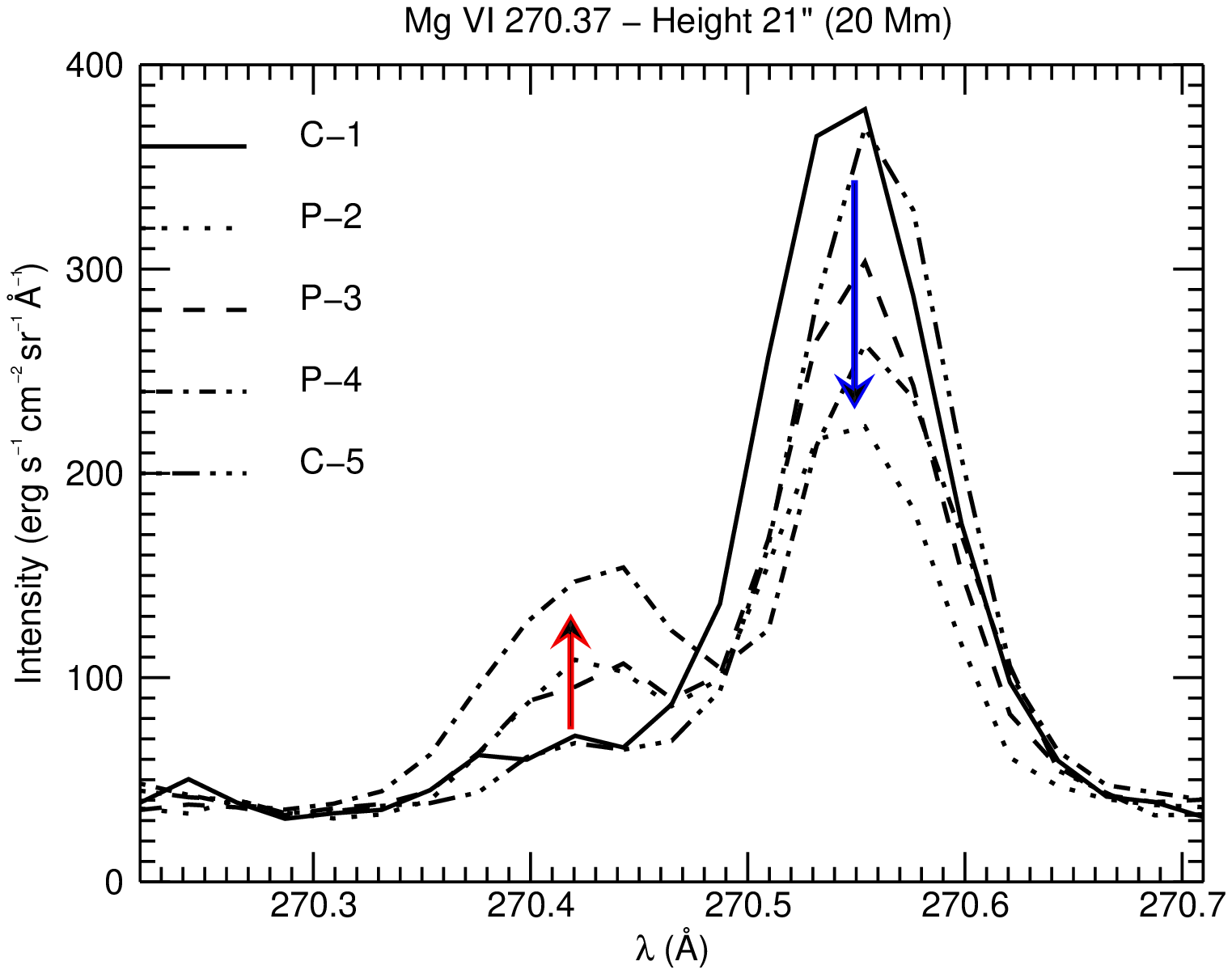}}
    \resizebox{\hsize}{!}{\includegraphics{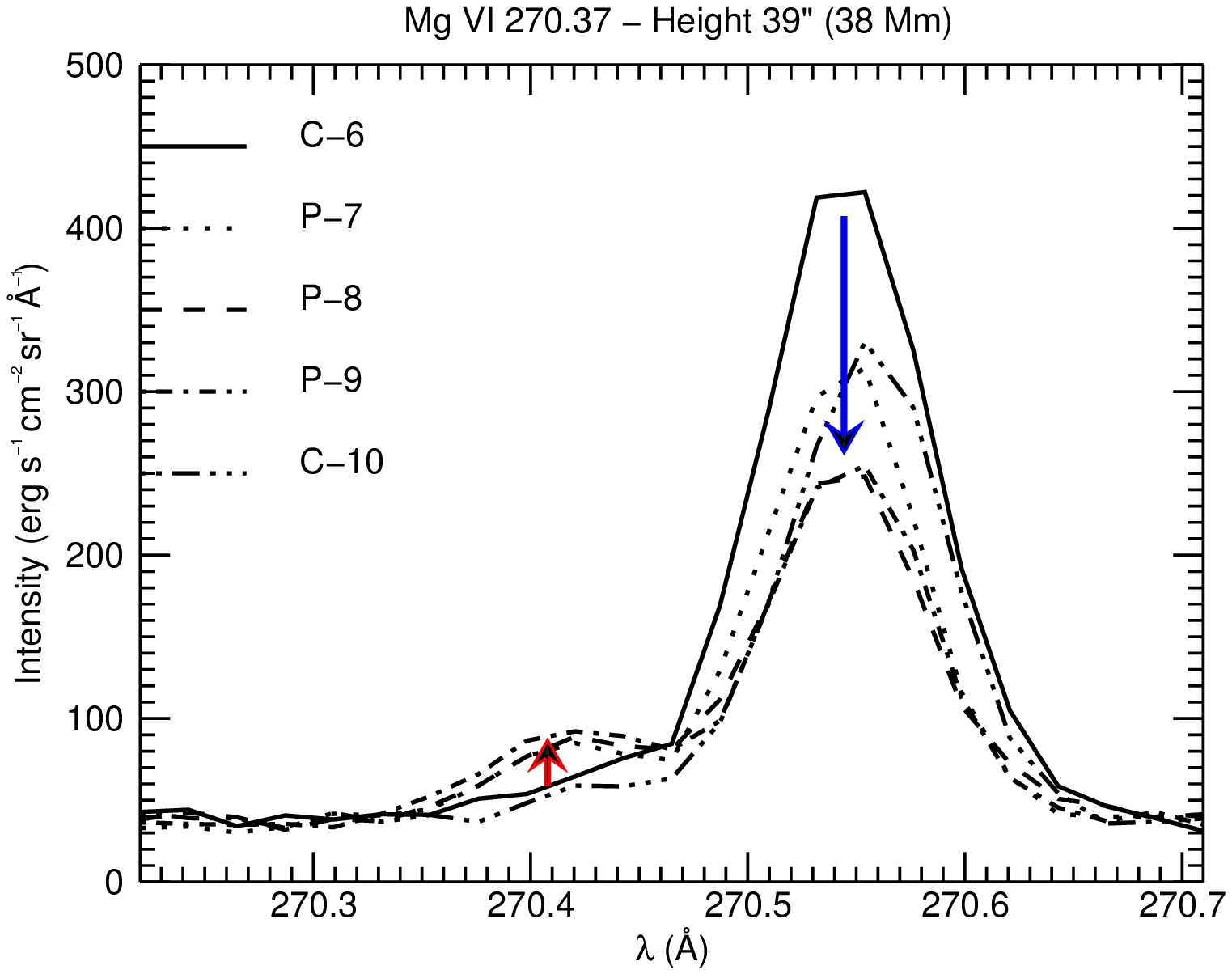}}
    \caption{{Same as Fig.~\ref{prof192} for} \ion{Mg}{vi} and \ion{Fe}{xiv} line profiles (blend). Upward and downward arrows indicate an increase in line emission for the \ion{Mg}{vi} line, and the effect of absorption and emissivity blocking due to the prominence in the \ion{Fe}{xiv} line.}
    \label{prof270}
\end{figure}

The  \ion{Mg}{vi} 270.40 raster image shows dark features at the location of the PCTR emission region.
This is due to the blend with the coronal \ion{Fe}{xiv} line at 270.52~\AA\ ($\log(T)=6.3$).
This is clearly seen when looking at the averaged line profiles
presented in Fig.~\ref{prof270}. 
In fact, it is possible to separate
these two lines by fitting Gaussian profiles. We find that the
prominence is then seen in (weak) emission in the \ion{Mg}{vi} line
and in absorption in the coronal line of \ion{Fe}{xiv}, as
expected. 

The line profiles on Fig.~\ref{prof270} show that at the lowest altitude in the prominence, the level of emission of the \ion{Mg}{vi} line is not negligible but is still, however, less than the contribution of
\ion{Fe}{xiv} 270.52~\AA. On the other hand, at greater height, the prominence is fainter, and the
emission in the \ion{Fe}{xiv} line is high, while negligible in the
\ion{Mg}{vi} line. Therefore, the image we see in the \ion{Mg}{vi}
raster is the result of the integration along the line of sight of
the coronal emission in the \ion{Fe}{xiv} line, with an additional,
small contribution from the \ion{Mg}{vi} 270.40~\AA\ line.

\section{\ion{He}{ii} line and continuum }\label{s:he2-line}

\begin{figure}
\centering
\resizebox{\hsize}{!}{\includegraphics{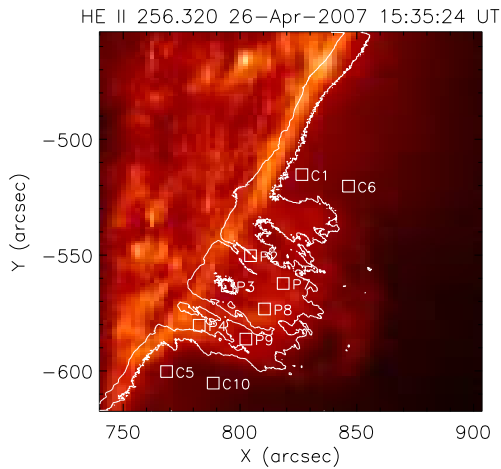}}
\caption{EIS raster in the 256~\AA\ window with \Ha\ contours from SOT.}
\label{HeIIraster}
\end{figure}

\begin{figure}
    \centering
    \resizebox{\hsize}{!}{\includegraphics{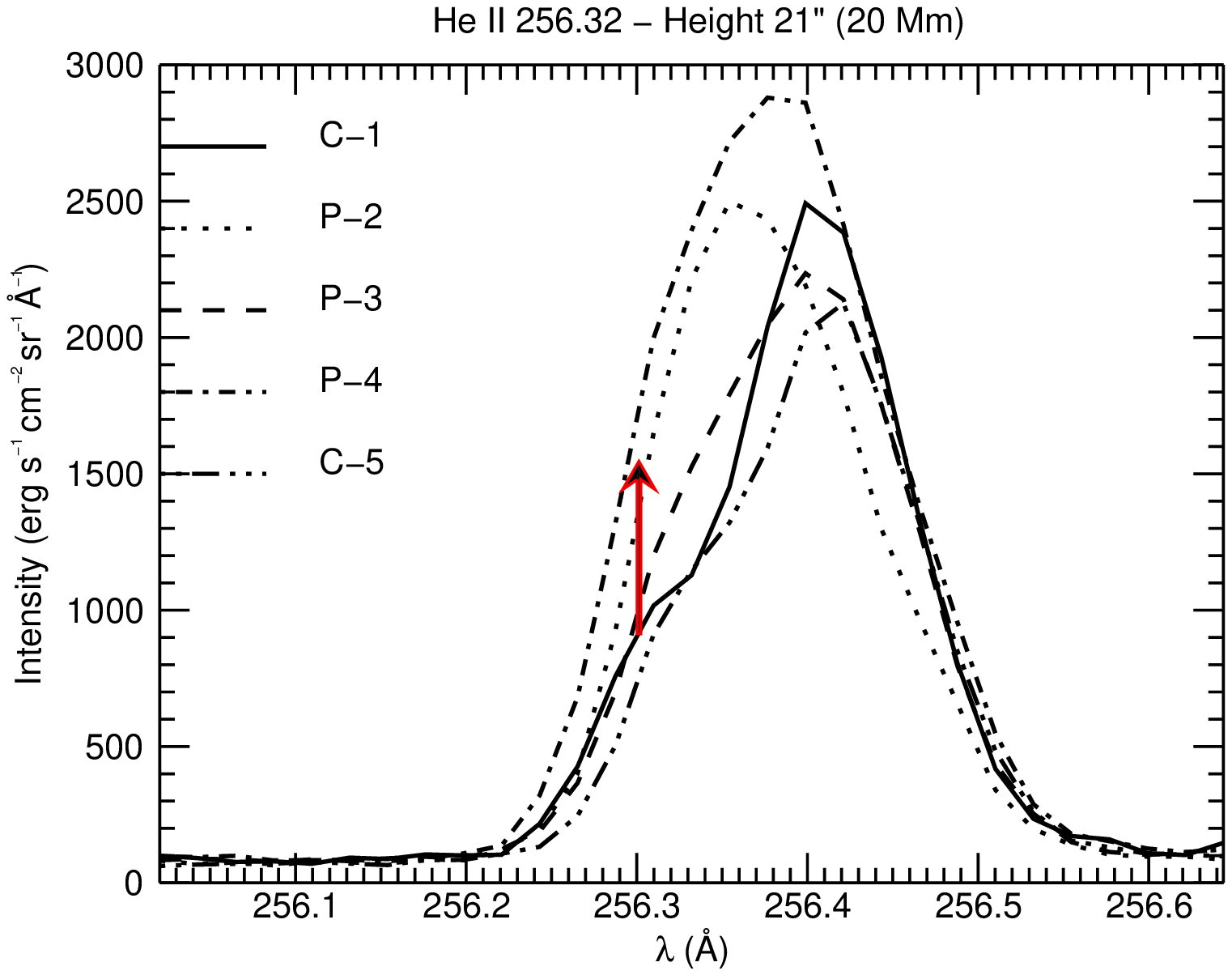}}
    \resizebox{\hsize}{!}{\includegraphics{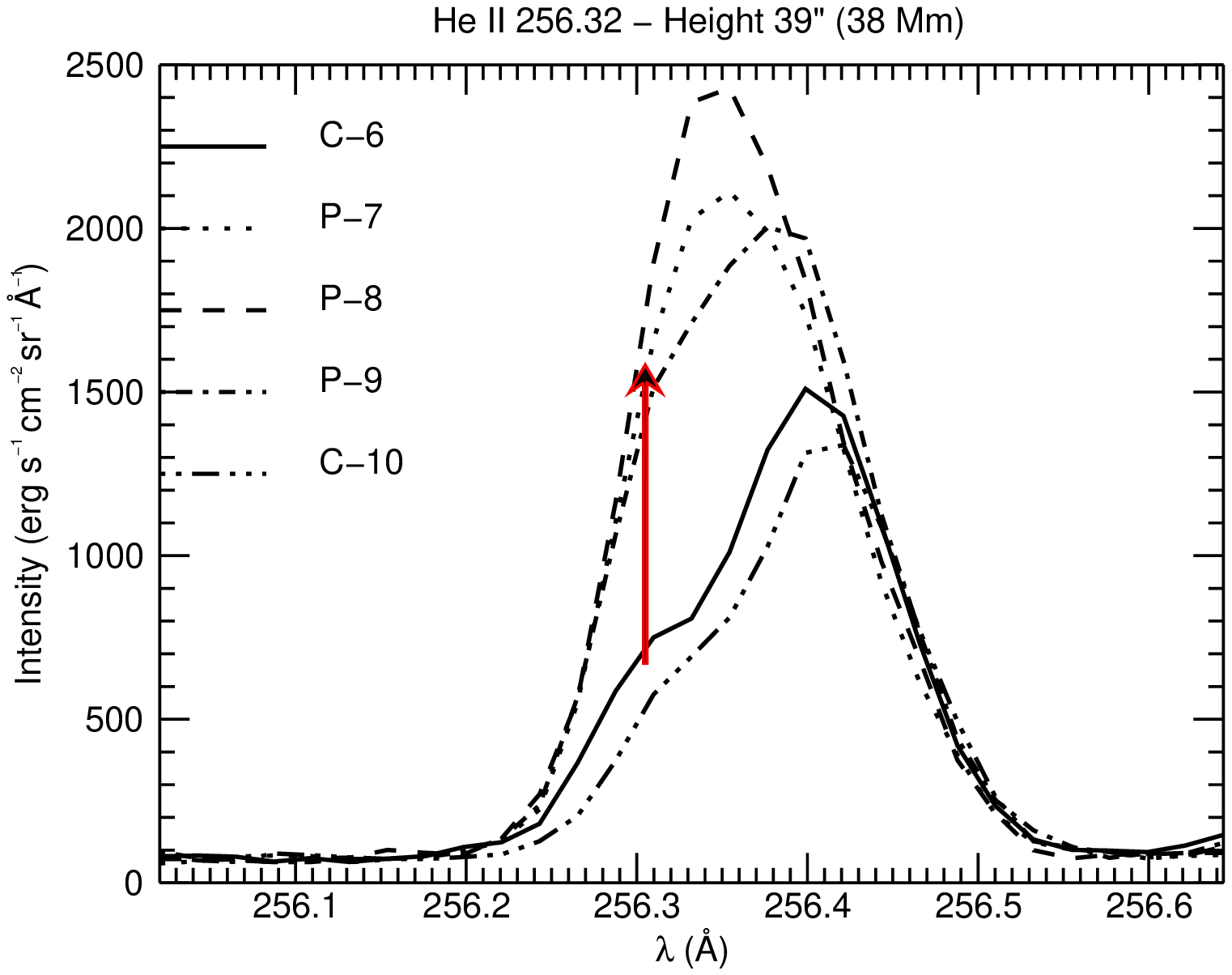}}
    \caption{{Same as Fig.~\ref{prof192} for} \ion{He}{ii} 256 line profiles (blend). Upward arrows indicate an increase in line emission in the blue wing due to the prominence.}
    \label{prof256}
\end{figure}

The EIS raster in the 256~\AA\ spectral window is shown in Fig.~\ref{HeIIraster}.
with superimposed contours from the \Ha\ intensity map obtained by SOT.
However, interpretation of the raster images is not straightforward, as a number of blends (of a coronal origin) with the \ion{He}{ii} line are present (see Table~\ref{tab:eislines} and the line profiles in Fig.~\ref{prof256}).

\subsection{\ion{He}{ii} 256 \AA\ line deblended}
\label{s:deblend}

The \ion{He}{ii} 256.32~\AA\ line is the Lyman~$\beta$ resonance line of ionized helium. It is the only 'cool' line observed by EIS. This line, together with the Lyman $\alpha$ line of \ion{He}{ii} at 304~\AA\ {and other \ion{He}{ii} lines}, is important for prominence diagnostics, and namely for the determination of the helium ionization degree and the study of the PCTR {\citep[e.g.][]{1979ApJ...232..929M,2001A&A...380..323L,2004ApJ...617..614L}}. In order to obtain pure \ion{He}{ii} 256 line emission originating in the prominence itself, we have to {separate} it {from} the \ion{He}{ii} line emission of coronal origin and from instrumental scattering, and {from} the coronal line blends. All these components together are here referred to as the {256-blend}. 

In this paper we do not intend to reconstruct the true \ion{He}{ii} line profiles in the prominence, but rather we estimate realistic limits for the integrated line intensities. These are then compared directly with typical model predictions. In order to avoid problems with coronal brightening close to the limb, and also to compare with the measurements of other authors of the off-limb coronal spectra, we choose the  height $H_2=38000$~km.

To deblend the coronal {256-blend}, we follow the procedure of P. Young (private communication, see also the EISWiki\footnote{\url{http://msslxr.mssl.ucl.ac.uk:8080/eiswiki/Wiki.jsp?page=HeIIOffLimb}, accessed April 2010.}) to disentangle between the coronal \ion{He}{ii} component and coronal blends. He used two Gaussians to fit the whole coronal {256-blend}. The first one at 256.321~\AA\ gives integrated line intensity $E=37$ (in cgs units), while the other at 256.413 \AA\ gives $E=139$. The total intensity is thus equal to 176. Therefore, for his EIS coronal observations, the ratio between the two Gaussian components is 0.27. We did the same for our coronal profiles at boxes C-6 and C-10 and found a ratio around 0.2, similar to Young's value. For coronal blends, Young (EISWiki) gives the ratios of individual blending lines to the total intensity of blends. Similar results were also obtained by E. Landi (private communication), who has computed the coronal blends using CHIANTI and the DEM model of the quiet solar corona recently derived from EIS data by Landi \& Young (2010). The blending lines are listed in Table~\ref{tab:eislines}. We are particularly interested by the \ion{Si}{x} line, which is the strongest blend. Part of its background coronal emission can be absorbed by the prominence due to expected high \ion{He} {ii} line opacity \citep[\ion{He}{ii} Lyman $\beta$ is optically thick as predicted by non-LTE prominence models of][]{2001A&A...380..323L}.

All coronal blends are attenuated at the position of the prominence due to (a) absorption by \ion{H}{i} and \ion{He}{i} resonance continua, and (b) emissivity blocking by the prominence and its surrounding cavity \citep{2005ApJ...622..714A,2008ApJ...686.1383H}. While the emissivity blocking is about the same for similar coronal ions (see Sect.~\ref{s:blocking}), the absorption due to photoionization by resonance continua is more complex, depending on the line wavelength and on the ionization state of the hydrogen-helium plasma. 

The only pure (in the sense that it is not blended by cooler lines) coronal line in our data set is the \ion{Fe}{xii} 195.12~\AA\ line. However, the radiation in this line is absorbed also by the \ion{He}{ii} resonance continuum which further complicates the situation. 
In order to see whether this \ion{Fe}{xii} line attenuation, which we directly measure in our EIS data set, can be used to estimate the attenuation of the {256-blend}, we have computed theoretical attenuation factors at 195~\AA\ and 256~\AA\ using Eq.~(11) and Table~2 of \cite{2005ApJ...622..714A}. Note that the opacities of the \ion{H}{i} and \ion{He}{ii} resonance continua scale as $\lambda^3$, while that of \ion{He}{i} scales as $\lambda^2$. For a reasonable range of optical thicknesses at these two wavelengths, we derive the corresponding attenuation factors $\frac{1}{2}(1+\exp(-\tau))$. 
For this we use the same ionization degrees as in Table~3 of \cite{2005ApJ...622..714A}. Finally, we arrive at the conclusion that the
attenuation factors are quite similar for both wavelengths of
interest, the maximum difference being around 10\% for $i=0.5$,
$j_1=0.3$, and $j_2=0$ \citep[ionization degree of hydrogen, neutral and ionized helium, respectively -- see notation in][]{2005ApJ...622..714A}.

Therefore, for 256~\AA\ we use the same total attenuation as we can derive from our 195~\AA\ line observations. The
latter is the product of $\frac{1}{2}(1+\exp(-\tau))$ and the coronal
emissivity-blocking factor $r_b$. By this factor we multiply the coronal
blends at boxes C-6 and C-10, for each prominence position. Although
the coronal intensities at C-6 and C-10 differ (the latter being lower
due to the presence of the cavity), our consistent treatment of the
emissivity blocking gives similar results using both boxes, to
within 5\% difference. Therefore we take the average values obtained
separately for boxes C-6 and C-10. From the total intensity
(attenuated) of all coronal blends we extract the \ion{Si}{x} intensity
using the ratios given by Young (EISWiki).

To {estimate the} pure \ion{He}{ii} integrated line emission in prominence positions P-7, P-8 and
P-9, we use the following four approximate approaches:

\paragraph{Model 1:}
From the total integrated intensity at the prominence position we first subtract the
full coronal \ion{He}{ii}, assuming that it is due to an instrumental
scattering of the strong disk radiation. Then we also subtract all
coronal blends properly attenuated.

\paragraph{Model 2:}
Same as Model 1, but we add back one half of \ion{Si}{x} blend,
assuming that roughly one half (the radiation from behind the prominence) was fully absorbed by the prominence
\ion{He}{ii} line opacity.

\paragraph{Model 3:}
Same as Model 1, but we subtract only one half of the
coronal \ion{He}{ii} component. Here we assume that it is of truly coronal
origin and thus one half of it is fully absorbed by the prominence
\ion{He}{ii} line opacity.

\paragraph{Model 4:}
As Model 3, but we add back one half of \ion{Si}{x} blend, assuming
that roughly one half was fully absorbed by the prominence \ion{He}{ii} line
opacity.

\medskip 

These four models are aimed at estimating some limiting values of the
\ion{He}{ii} line intensities in the prominence, and thus we ignore some
secondary factors. Since we do not reconstruct the true line
profiles here, we cannot compute realistic absorption of the
background coronal radiation by the \ion{He}{ii} line in the prominence,
which depends on the relative width and shift of the coronal
emission lines and \ion{He}{ii} line absorption profile. We thus consider
only two limiting cases, i.e. full absorption or no absorption of the \ion{Si}{x} line radiation. 
We also don't disentangle between the continuum absorption and emissivity blocking factors {(while the two processes were disentangled in Sect.~\ref{s:abs}),} and take into account only the combined factor from our 195~\AA\ measurements. 
To be fully consistent, in
Model 2 one should subtract the foreground \ion{Si}{x} radiation attenuated only
by the emissivity blocking factor, while in Model 3 this factor
should be used to attenuate one half of the presumably coronal \ion{He}{ii}
radiation (the background part being fully absorbed by the prominence).

Our results for the four models and three prominence boxes are
summarized in Table~\ref{tab:fit256}. 
\begin{table*}
\caption{Integrated intensities for the \ion{He}{ii} 256~\AA\ line and the blending lines  in  different regions of the corona (C) and the prominence (P).}
\label{tab:fit256}
\centering
\begin{tabular}{ccccccccc}
\hline\hline
Integrated Intensity / Box & P-7& P-8& P-9& C-6& C-10&LY(obser)& LY (pred) & Y (obser)\\
\hline
Total emission  & 323 & 353 & 328 & 224 & 191 & &   & 176\\
\ion{He}{ii}      &    &   &   & 50 & 40 &&   &37\\
coronal blends    & 135 & 125 & 138 & 174 & 151 & & 157 & 139\\
\ion{Si}{x} 256.37 & 92 & 85 & 94 &  &&&101&95\\
1/2 (\ion{Si}{x} 256.37) & 46 & 42 & 47 &&&&\\
\ion{Si}{x} 261.04 &&&&&&104&90&84\\
\hline \ion{Fe}{xii} 195.12 & 520 & 482 & 534 & 681 & 575 & 581 & 761 & \\
\hline
\ion{He}{ii} (model 1) & 143 & 183 & 145 &&&&&\\
\ion{He}{ii} (model 2) & 189 & 225 & 192 &&&&&\\
\ion{He}{ii} (model 3) & 165 & 205 & 167 &&&\\
\ion{He}{ii} (model 4) & 211 & 247 & 214 &&&\\
\hline \hline
\end{tabular}
\tablefoot{The total emission quoted here is corrected from the background emission, assumed to be of instrumental origin. LY refers to the paper of \cite{2010ApJ...714..636L} and to CHIANTI computations by E. Landi (private communication), (obser) to observed values, (pred) to predicted values, Y to Young's EISWiki. Note that LY values correspond to a height in the corona of about $25000$~km, somewhat {lower than the present work}.}
\end{table*}
{In this table} the blends have been attenuated and the prominence \ion{He}{ii} line emission estimated using our four models for $H=38 000$~km. 
We see that the integrated intensity values are  all in a narrow range: 143 to 247~erg s$^{-1}$ cm$^{-2}$ sr$^{-1}$. {As we will see in Sect.~\ref{s:nlte},} these values can be compared with the non-LTE modelling of \cite{2001A&A...380..323L} and \cite{2004ApJ...617..614L}. They also represent constraints for new 2D models \citep{2008A&A...490..307G,2009A&A...503..663G,2009A&A...498..869L}.

In a next paper, we plan to follow these guidelines and restore the true \ion{He}{ii} line profiles in the prominence, but also in the coronal boxes. One problem is that using two Gaussians for the decomposition of the coronal {256-blend} is a rather crude approximation. We made some preliminary tests with the detailed profile fitting and got a rather asymmetrical and blue-shifted \ion{He}{ii} coronal component, while the coronal blends also produce a non-Gaussian profile due to a slight relative wavelength shift of individual blending lines. Therefore, the detailed profile fitting can lead to a somewhat modified decomposition of the coronal profile.
Moreover, having a more realistic \ion{He}{ii} coronal component profile, we could decide about its origin. We are able to estimate theoretically the amount of \ion{He}{ii} coronal radiation which is due to scattering of the disk radiation by coronal \ion{He}{ii} ions. We are also able to compare the \ion{He}{ii} coronal component with the shape of the disk emission profile (the raster extends to the disk), and see whether this component resembles the disk one. This would then mean that it could be due to instrumental scattering.
Finally, Young (EISWiki) has pointed out that this component of the 256~\AA\ coronal blend could be related to a so far unidentified solar coronal blend(s). In such a case, the de-blending procedure will be the same as when we consider this component to be \ion{He}{ii} of coronal origin. We postpone such analysis to our next paper.

\subsection{He II continuum}

The continuum window between 189~\AA\ and 210~\AA\ had been selected to investigate the level of emission in the helium resonance continua, and namely \ion{He}{ii} continuum with the edge at 228~\AA. However, the emission in this continuum {below 210~\AA} is too low in the prominence and no signal can be reliably measured.
{Using OSO-7 observations, \cite{1976ApJ...203..509L} measured the intensity at the edge of the \ion{He}{ii} continuum in a prominence to be a few times $10^{-13}$~erg s$^{-1}$ cm$^{-2}$ sr$^{-1}$ Hz$^{-1}$. This is below what EIS can detect in a prominence at $\lambda < 210$~\AA. \cite{2007PASJ...59S.727Y} observed \ion{He}{ii} enhanced continuum emission in an active region transition region brightening at a level of $\sim 10^{-11}$~erg s$^{-1}$ cm$^{-2}$ sr$^{-1}$ Hz$^{-1}$.}

\subsection{Non-LTE radiative transfer calculations}\label{s:nlte}

As a preliminary analysis, we use a grid of non-LTE models computed by \cite{2004ApJ...617..614L} to identify the formation mechanisms of the \ion{He}{ii} 256~\AA\ line in the prominence \citep[also discussed in][]{2009A&A...503..663G}. 
{In the study by \cite{2004ApJ...617..614L}, the prominence is represented as a 1D plane-parallel slab standing vertically above the solar surface. Inside the prominence, the pressure and the temperature vary along the line-of-sight according to the theoretical formulae of \cite{1999A&A...349..974A}. The pressure profile is derived from the magneto-hydrostatic equilibrium, while the temperature profile is purely empirical. These prominence models include a prominence-to-corona transition region.}
By using {these} simple prominence models, we are able to reproduce the range of inferred integrated intensities of the \ion{He}{ii} 256~\AA\ line reported in Table~\ref{tab:fit256}. Here we only compare theoretical and observed intensities in that line. As an example, Table~\ref{tab:model} presents a typical model yielding an integrated intensity in the \ion{He}{ii} line {of $200$~erg s$^{-1}$ cm$^{-2}$ sr$^{-1}$,} comparable to the inferred {value}.
\begin{table}
    \caption{Physical parameters for a non-LTE radiative transfer model.}
        \label{tab:model}
        \centering
        \begin{tabular}{lc}
                \hline\hline
                Parameter & Value\\
                \hline
                Central temperature	& $8700$~K\\
                Surface temperature	& $96500$~K\\
                Central pressure	& $0.33$~dyn cm$^{-2}$\\
                Surface pressure	& $0.22$~dyn cm$^{-2}$\\
                Column mass			& $2.5\times 10^{-4}$~g cm$^{-2}$\\
                {Hydrogen column density} & $10^{20}$~cm$^{-2}$\\
                \hline\hline
                Intensity of the 256~\AA\ line & $200$~erg s$^{-1}$ cm$^{-2}$ sr$^{-1}$\\
                \hline
        \end{tabular}
\end{table}

From the result of our non-LTE radiative transfer calculations, we estimate that about 90\%
of the prominence emission in the \ion{He}{ii} line at 256~\AA\ close to the prominence-corona boundary comes from the scattering of
the incident radiation, while the rest of the emission is due to other atomic processes (mainly
 collisional excitations). This fraction decreases to about 35\% in the central parts of the prominence. Most of the line emission comes from the PCTR (where the \ion{He}{ii} ions are mostly present) and is due to resonant scattering of the incident radiation.
{This is consistent with the fact that the \Ha\ and \ion{He}{ii} 256 lines are not always correlated (as seen in Fig.~\ref{sot}). When the 256~\AA\ line is seen with no \Ha\ counterpart emission, we probably see the PCTR itself (the LOS does not intercept the central parts of the prominence). In addition, if most of the \ion{He}{ii} emission in this line results from the scattering of the incident radiation, we do not expect a strong correlation with other lines emitted in the PCTR by thermal processes.}
{It is worth noting that our prominence model has a maximum temperature of $10^5$~K at the coronal boundary, so it is not suitable for an analysis of the other PCTR lines observed by EIS.}

Using these results, we have an estimate of the continuum intensity of the hydrogen and helium continua. The predicted \ion{He}{ii} continuum intensity is lower than $10^{-13}$ erg s$^{-1}$ cm$^{-2}$ sr$^{-1}$ Hz$^{-1}$ in the spectral range observed by EIS with the CCD~B. 

Because of the uncertainties attached to the determination of the observed integrated intensity of the \ion{He}{ii} line, there is some uncertainty regarding the uniqueness of the model presented in Table~\ref{tab:model}. In particular, the central pressure seems large. It is possible that we could obtain a different model with a lower pressure if we took a different temperature profile. On the other hand, we may be seeing the prominence under an angle such that the line-of-sight is more or less along the field lines, a situation that our simple 1D slab model cannot reproduce very well. 
{This is related to the fact that the value of the hydrogen column density indicated by our non-LTE model fitting of the intensity of the \ion{He}{ii} line alone seems larger than expected from the relation between the optical depth at 195~\AA\ and the hydrogen column density given by Equation~12 in  \cite{2008ApJ...686.1383H}. With the values for $\tau_{195}$ reported in Table~\ref{tab:percent}, that relation gives us a hydrogen column density of $n_\mathrm{H} \sim 10^{17}-10^{18}$~cm$^{-2}$, which is about two orders of magnitude lower than what we obtain from the model of Table~\ref{tab:model}. Similarly, the hydrogen column density quoted in Table~\ref{tab:model} yields $\tau_{195}\sim16$, a value too large compared to what we derived from the observations (Table~\ref{tab:percent}).}
{Note that, based on the modelling developed by \cite{2008A&A...490..307G}, \cite{berlicki} were able to construct a new model allowing them to reproduce the observed profiles and line intensities of hydrogen Lyman and \Ha\ lines observed in the same prominence with 40 threads of temperature 8000~K, central pressure 0.035~dyn cm$^{-2}$, and a column mass $3\times 10^{-6}$~g cm$^{-2}$ across the field lines. With 40 threads, this yields a total column mass of $1.2\times 10^{-4}$~g cm$^{-2}$ at the centre of the prominence, only a factor two less than our value (Table~\ref{tab:model}).}

{The fact that we obtain different results in our modelling is not very surprising. Berlicki et al. used a complete set of hydrogen lines to find a 2D model which reproduces well their observed line profiles. In the present study, our 1D model is simply based on the estimated integrated intensity of the \ion{He}{ii} line at 256~\AA. If anything, this highlights the need to consistently use the information brought by both hydrogen and helium line profiles. We have made some preliminary attempts in combining both types of models in order to compute the intensity of the \ion{He}{ii} 256 line using the Berlicki et al. models, but the resulting integrated intensity is too low. This may be a signature of new diagnostics of the PCTR provided by the \ion{He}{ii} lines.}

{We also stress that this prominence is different from the prominence observed the previous day and analysed in \cite{2008ApJ...686.1383H}. In particular, we do not see the same absorption by the prominence in EIS images  because there are coronal structures in the foreground. STEREO A and B show effectively coronal structures in 171~\AA\ projected inside the cavity. This in turn may affect our estimates of the optical depth at 195~\AA.}
 
In the next paper, we will attempt to recover the full line profile of the \ion{He}{ii} 256~\AA\ line from the EIS data in the prominence, and we will make a more detailed quantitative comparison with predicted line profiles.
{We will also study the other He lines predicted by our models.}

\section{Conclusions}

It is shown that the EUV Imaging Spectrometer of Hinode is a powerful instrument which is bringing new insights into the physics of solar prominences, and in particular of the prominence-to-corona transition region. Despite the fact that EIS sensitivity is {tuned} towards short wavelengths bands, there are several lines which are suitable to investigate the properties of the prominence region and the surrounding corona in terms of temperature and density.

In particular, the \ion{He}{ii} 256.32~\AA\ line is a strong emission line showing details about the prominence fine structure.
Transition region lines such as \ion{Fe}{viii} 185.21~\AA, \ion{Fe}{viii} 186.60~\AA\ and \ion{Si}{vii} 275.35~\AA\ are also well-observed in the prominence, while coronal lines clearly show the prominence as a dark feature. {This is} due to absorption of the coronal EUV radiation by the resonance continua of hydrogen and helium, and by the emissivity blocking effect.
{We have used the intensities observed by EIS at 195~\AA\ and by XRT to estimate the optical depth at 195~\AA\ in the prominence.}

Although the \ion{He}{ii} 256.32~\AA\ line is listed  with a temperature of formation of about 50\,000~K in \citet{1998A&AS..133..403M}, this value should not be taken as representative of the prominence plasma seen in Fig.~\ref{rasters}.
The prominence plasma is optically thick {at this wavelength}, and there is an important contribution of resonant scattering of the incident radiation coming from the solar disk and the surrounding corona. This situation is similar to the case of the first resonance line of ionised helium at 303.78~\AA\ \citep{2001A&A...380..323L,2008AnGeo..26.2961L}. This provides constraints to new 2D models \citep[see][]{2008A&A...490..307G,2009A&A...503..663G,2009A&A...498..869L}.

Off-limb raster images from EIS must be looked at carefully in order to separate the contributions of the different lines blended with the nominal line. The interpretation of the \ion{He}{ii} raster image is not straightforward. In this paper we have analyzed the blend of the \ion{He}{ii} 256~\AA\ line and obtained a reasonable estimate of the \ion{He}{ii} line integrated intensities in several prominence locations.
These are consistent with results of non-LTE modelling \citep{2004ApJ...617..614L}. 
{However, the corresponding theoretical hydrogen column density is about two orders of magnitude higher than what is inferred from the opacity estimates at 195~\AA.}

The detailed analysis of the line profiles of the {256-blend} will be studied in a next paper. {We will also address the question of the diagnostics obtained from the 256~\AA\ line \textit{vs.} the 304~\AA\ line of \ion{He}{ii}. We do not expect to find a strong correlation between the two lines in our theoretical results. Observational studies \citep[e.g.,][]{1978ApJ...222..707G} found that both lines were uncorrelated in the Quiet Sun. We will also investigate whether the \ion{He}{ii} 256~\AA\ line could be used in conjunction with other lines to estimate the \ion{He}{ii}/\ion{He}{i} abundance. }

The spectral diagnostics of the filaments and prominences observed for JOP 178 will benefit from the large number of lines observed by the three space-based spectrometers EIS, CDS, and SUMER, from the ground-based \Ha\ data, and from our non-LTE radiative transfer codes to analyse the hydrogen \citep{2000SoPh..196..349G,2007A&A...472..929G} and helium \citep{2001A&A...380..323L,2004ApJ...617..614L} spectra.
We plan to combine these observations to further constrain our non-LTE models and infer the thermodynamical properties of the prominences and filaments and their surroundings.

\begin{acknowledgements}
{We thank the referee for their constructive comments.}
We are grateful to Giulio Del Zanna, Stanislav Gun\`ar, Pierre Gouttebroze, Enrico Landi, and Peter Young for helpful discussions, as well as to our colleagues from the International Teams 123 and 174 of the International Space Science Institute (ISSI) in Bern.
The support of ISSI is acknowledged.
NL  was partially supported by the European Commission through the SOLAIRE Network (MTRN-CT-2006-035484) and by STFC Rolling Grant (ST/F002637/1). 
PH was partially supported by the ESA-PECS project No. 98030. He also acknowledges the hospitality and support of the Paris Observatory.
We thank Pavol Schwartz for helping us with Fig.~\ref{cuts}.
Hinode is a Japanese mission developed and launched by ISAS/JAXA, with NAOJ as domestic partner and NASA and STFC (UK) as international partners. It is operated by these agencies in co-operation with ESA and NSC (Norway). 
This research has made use of NASA's Astrophysics Data System.
\end{acknowledgements}

\bibliographystyle{aa}
\bibliography{../../../../LATEX/all2}

\end{document}